\documentclass[aps,twocolumn,superscriptaddress,10pt]{revtex4-2}

\usepackage{graphicx}
\usepackage{amsmath,amssymb,amsfonts}
\usepackage{bm}
\usepackage{xcolor}
\usepackage[caption=false]{subfig}
\usepackage{hyperref}
\usepackage{lineno}
\usepackage{etoolbox}
\AtBeginEnvironment{figure}{\nolinenumbers}
\AtBeginEnvironment{figure*}{\nolinenumbers}

\begin{document}

\title{Active normal-fluid feedback in quantum turbulence from holography}

\author{Yu-Ping An}
\email{yuping.an@campus.technion.ac.il}
\affiliation{Department of Physics, Technion, Haifa 32000, Israel}

\author{Li Li}
\email{liliphy@itp.ac.cn}
\affiliation{Institute of Theoretical Physics, Chinese Academy of Sciences, Beijing 100190, China}
\affiliation{School of Physical Sciences, University of Chinese Academy of Sciences, Beijing 100049, China}
\affiliation{School of Fundamental Physics and Mathematical Sciences, Hangzhou Institute for Advanced Study, University of Chinese Academy of Sciences, Hangzhou 310024, China}

\author{Makoto Tsubota}
\email{tsubota@omu.ac.jp}
\affiliation{Department of Physics, Nambu Yoichiro Institute of Theoretical and Experimental Physics (NITEP), Osaka Metropolitan University, 3-3-138 Sugimoto, Osaka 558-8585, Japan}

\author{Sebastian Waeber}
\email{waeber@post.bgu.ac.il}
\affiliation{Department of Physics, Ben-Gurion University of the Negev, David Ben Gurion Boulevard 1, Beer Sheva 84105, Israel}


\author{Hua-Bi Zeng}
\email{zenghuabi@hainanu.edu.cn}
\affiliation{Center for Theoretical Physics, Hainan University, Haikou 570228, China}

\begin{abstract}
The two-fluid model provides a powerful phenomenological framework for superfluids, where components interact through vortex-induced mutual friction. Quantum turbulence in such coupled two-fluid systems has become a central theme in low-temperature physics. However, the fundamental incompatibility between discrete vortex-line and continuum descriptions has historically forced reliance on phenomenological parameters. Here, we present an ab initio holographic simulation of this active two-fluid dynamics. Using a fully backreacted gravitational model, we intrinsically incorporate reciprocal momentum exchange and finite-temperature dissipation without empirical inputs. Contrasting continuously driven and freely decaying turbulence, we uncover pronounced non-Gaussian tails and spatial anisotropy in the normal-fluid velocity statistics, arising directly from the dynamical quantized vortex backreaction. Raising the temperature suppresses normal-fluid velocity fluctuations despite increasing the vortex density. This suppression is unambiguously traced to the thermal depletion of the superfluid condensate. Our results establish holographic duality as a rigorous first-principles framework for quantum turbulence, revealing the active role of the normal fluid far from equilibrium and yielding predictions for ongoing cold-atom and superfluid experiments.
\end{abstract}
\maketitle

Superfluidity stands as one of the most profound manifestations of macroscopic quantum coherence. At finite temperatures, its phenomenology is elegantly captured by Landau's two-fluid model, which partitions the system into an inviscid superfluid component and a viscous normal fluid component~\cite{tisza1938transport,landau41,khalatnikov2018introduction}. For decades, a prevailing theoretical simplification has treated the normal fluid as a passive thermal bath that merely provides a dissipative backdrop for the quantized vortices in the superfluid~\cite{schwarz85,schwarz88,adachi10}. However, this passive picture was fundamentally challenged by recent groundbreaking flow-visualization experiments in superfluid $^4$He~\cite{marakov2015visualization,gao2017energy,mastracci2018exploration,mastracci2019particle}. These experiments directly observed that the initially laminar normal-fluid profile becomes distorted and transitions into a fully turbulent state solely due to the dynamical feedback exerted by the quantized vortex tangle~\cite{marakov2015visualization}. The normal fluid, far from being a passive sink, exhibits highly non-Gaussian velocity statistics and actively exchanges energy and momentum with the superfluid on an equal footing~\cite{mastracci2018exploration}.

Capturing this active, far-from-equilibrium two-way coupling from first principles presents a formidable theoretical bottleneck. Conventional approach couples the Lagrangian Vortex Filament Model (VFM) for the superfluid with the Eulerian Navier-Stokes equations for the normal fluid~\cite{kivotides00,yui2020fully,kobayashi2024influence,galantucci20,tang2023imaging}. Despite its wide usage, this hybrid strategy suffers from a fundamental Lagrangian-Eulerian inconsistency. More critically, it inevitably relies on ad hoc phenomenological parameters such as the mutual friction coefficients~\cite{hall1956rotation,hall1956rotation2,barenghi1983friction} which must be calibrated and lack predictive power in the strongly interacting, far-from-equilibrium regimes relevant to experiments. Consequently, a genuinely first-principles framework that intrinsically unifies the vortex kinetics with macroscopic dissipative hydrodynamics has remained a central unfulfilled goal in quantum turbulence research~\cite{barenghi2023quantum,tsubota2025quantum}.

The holographic duality (AdS/CFT correspondence) offers a compelling framework by mapping strongly coupled quantum field theories to classical gravity in one higher dimension~\cite{zaanen2015holographic,hartnoll2018holographic}. It provides a first-principles description of dissipative hydrodynamics with fully consistent energy-momentum conservation, addressing precisely the Lagrangian-Eulerian inconsistency that plagues conventional hybrid models. Within this framework, the two-fluid structure emerges naturally from a single, self-consistent gravitational system, circumventing any dependence on phenomenological parameters~\cite{hartnoll2008building,herzog2009holographic,sonner2010gravity}. Nevertheless, previous holographic vortex studies have been almost exclusively restricted to the probe limit~\cite{chesler2013holographic,lan2016towards,wittmer2021vortex,lan2023heating,an2024interface,xia2026kibble,zeng2025dissipation}, where matter fields evolve on a fixed background, essentially imposing a one-way coupling where the normal fluid remains an infinite, unresponsive reservoir. This restricts the normal fluid to the very passive-bath assumption that experiments have overturned. 
Capturing the active two-way coupling demands a fully backreacted treatment, which comes at a formidable technical cost, namely 16 coupled, highly nonlinear PDEs without symmetry reduction, a barrier that has, until now, precluded first-principles simulations of two-fluid quantum turbulence.

\begin{figure*}[htbp]
\centering
\includegraphics[width=1\textwidth]{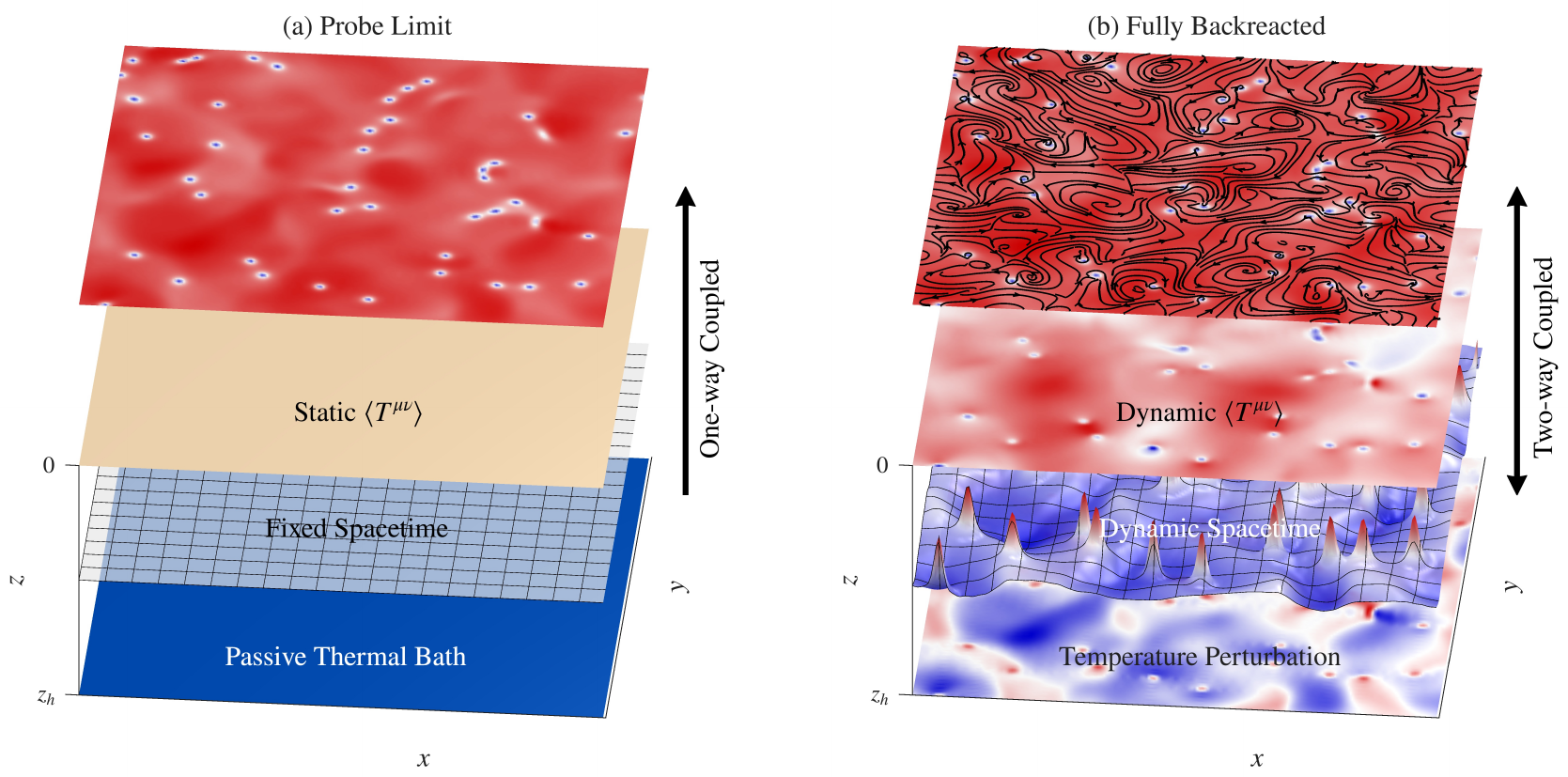}
\caption{
\textbf{Self-consistent holographic modeling of two-fluid dynamics.} The schematic contrasts the strictly one-way coupled probe limit (a) with the two-way coupled, fully backreacted model (b). \textbf{a}, In the probe limit, matter fields evolve on a rigid background, restricting the normal fluid to a passive thermal bath without dynamical vortex feedback. \textbf{b}, With full gravitational backreaction, a self-consistent two-way coupling emerges as bulk geometry responds to fluid fields. Top: superfluid condensate overlaid with normal-fluid velocity streamlines. Boundary ($z = 0$): metric expansion coefficients encode the stress-energy tensor $\langle T^{\mu\nu}\rangle$. Horizon ($z = z_h$): local temperature fluctuations reflect thermal dissipation and entropy production. The dynamical bulk (middle) intrinsically couples both components, generating mutual friction from first principles.
\label{fig:schematic}}
\end{figure*}

In this work, we overcome this long-standing limitation by performing the first real-time simulations of two-dimensional (2D) quantum turbulence using a fully backreacted holographic superfluid in a $(3+1)$-dimensional asymptotically Anti-de Sitter spacetime (AdS). 
Our setup dynamically solves the fully coupled Einstein-Maxwell-scalar equations, dynamically capturing continuous momentum transfer, finite-temperature dissipation, and mutual friction without any empirical input (see Fig.~\ref{fig:schematic}). Moreover, to establish a statistically steady turbulent state
we exploit the Landau instability~\cite{arean2024hydrodynamics} to continuously inject vortices homogeneously, avoiding the spatial inhomogeneities introduced by external stirring potentials. 

By systematically contrasting this continuously driven turbulent regime with freely decaying turbulent regime, we achieve a critical conceptual disentanglement: we rigorously isolate the genuine finite-temperature effects from the trivial effects of vortex density variations. Our simulations uncover two pivotal discoveries. First, the real-space velocity statistics exhibit pronounced non-Gaussian tails and spatial anisotropy that arise directly from the dynamical backreaction of quantized vortices onto the normal fluid, providing a direct macroscopic signature of the active normal-fluid response. Second, and most strikingly, we find that increasing the temperature suppresses the normal-fluid velocity fluctuations, even though it simultaneously increases the vortex number density. By contrasting this with the decaying case—where the vortex density drops while temperature stays nearly constant—we unambiguously demonstrate that the magnitude of normal-fluid fluctuations is not dictated by vortex density. Instead, this suppression is a genuine effect caused by the thermal depletion of the superfluid density, which inherently weakens the backreaction exerted by the vortices on the normal component.

Our results establish fully backreacted holographic superfluids as a rigorous, first-principles paradigm for quantum turbulence, completely bypassing the phenomenological parameters that limit conventional hybrid models. This framework not only elucidates the active role of the normal fluid far from equilibrium but also provides sharp, testable predictions for ongoing experiments on 2D cold-atom superfluids and superfluid films. The paper is organized as follows: we first detail the holographic setup and the extraction of two-fluid observables, then present the spectral and statistical analyses for both driven and decaying turbulence, and conclude by discussing the universality of our findings and their implications for higher-dimensional quantum turbulence.

\begin{figure*}[t]
    \centering
    \subfloat[]{\includegraphics[width=0.46\linewidth]{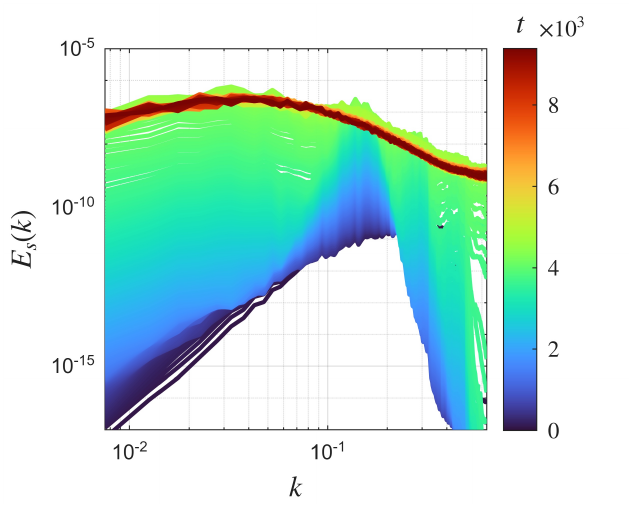}}
    \subfloat[]{\includegraphics[width=0.46\linewidth]{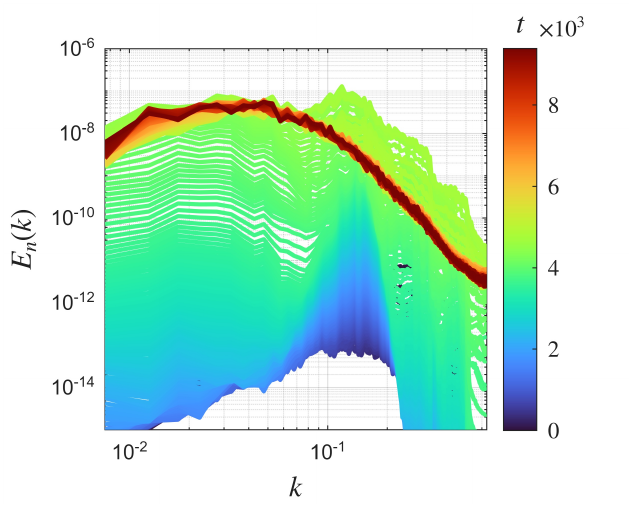}}\\
    \subfloat[]{\includegraphics[width=0.43\linewidth]{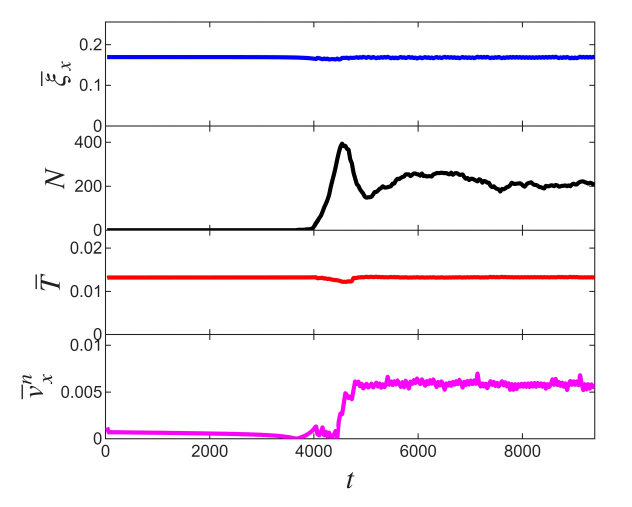}}
    \subfloat[]{\includegraphics[width=0.46\linewidth]{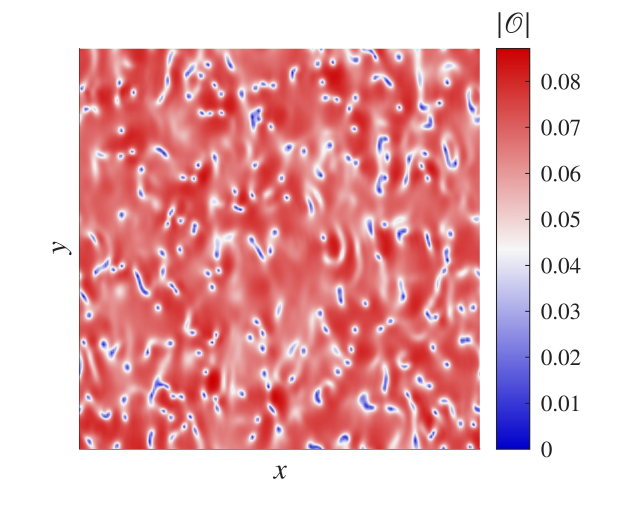}}
    \caption{\textbf{Demonstration of statistically steady state with $\bar T/T_c=0.635$ and $\bar\xi_x=0.17$.} \textbf{(a,b)} Time evolution of the superfluid and normal-fluid energy spectra, \(E_s(k)\) and \(E_n(k)\), showing relaxation to time-independent distributions after an initial transient. \textbf{(c)} Spatially averaged quantities—mean superfluid velocity \(\bar{\xi}_x\), vortex number \(N_v\), effective temperature \(\bar{T}\), and mean normal-fluid velocity \(\bar{v}_x^n\)—converge to plateau values with only small fluctuations. The vortex number stabilizes through sustained nucleation-annihilation balance. \textbf{(d)} Snapshot of the condensate magnitude \(|\mathcal{O}|\) at \(t \simeq 9000\), illustrating the vortex configuration in the steady state. Together, these panels confirm that the fully backreacted holographic framework generates and sustains 2D quantum turbulence from first principles, without empirical dissipation parameters.}
    \label{fig:tevol_driven}
\end{figure*}

\subsection*{Establishing a First-Principles Two-Fluid Turbulence}

The holographic dictionary maps the gravitational dynamics in the bulk to the real-time evolution of a strongly coupled quantum fluid on the boundary. Within this correspondence, the two-fluid structure is not imposed \textit{a priori} but emerges from the following action~\cite{hartnoll2008building,herzog2009holographic,sonner2010gravity}:
\begin{equation}
\label{eq:action}
\begin{aligned}
    S =&\frac{1}{16\pi G}\int d^4x \sqrt{-g}\left(R + \frac{6}{L^2}\right)\\&+\int d^4x \sqrt{-g}\left[-\frac{1}{4}F_{ab}F^{ab}- |D\Psi|^2- m^2 |\Psi|^2\right],
\end{aligned}
\end{equation}
where $G$ is the Newton constant, $R$ is the Ricci scalar, $L$ is the AdS radius, $F_{ab} = \partial_a A_b - \partial_b A_a$ is the $U(1)$ gauge field strength tensor, and $D_a \Psi = \nabla_a \Psi - iq A_a \Psi$ denotes the covariant derivative acting on the complex scalar field $\Psi$ with charge $q$ and mass $m$. The scalar field $\Psi$ condenses below a critical temperature $T_c$, spontaneously breaking the global $U(1)$ symmetry and giving rise to the superfluid order parameter $\langle \mathcal{O} \rangle$, while the dynamic black hole geometry dynamically encodes the viscous, dissipative normal fluid through the boundary stress-energy tensor $\langle T^{\mu\nu} \rangle$ and conserved current $\langle J^\mu \rangle$. 

Intuitively, the system's thermodynamics and hydrodynamics are encoded in the bulk geometry. The dynamic black hole embodies the thermalized, dissipative normal fluid, establishing its macroscopic temperature and generating its viscous transport through the irreversible infall of physical fields. Simultaneously, the complex scalar field provides the superfluid macroscopic wavefunction. As quantized vortices move, they locally warp the surrounding spacetime. Through fully backreacted Einstein equations, these dynamic deformations propagate to the boundary, manifesting as hydrodynamic drag. In this picture, two-way momentum exchange and mutual friction arise directly from first principles. This unified framework allows us to extract both velocity fields directly from the boundary data without auxiliary modelling: the superfluid velocity $\xi_i$ is obtained from the order parameter $\langle \mathcal{O} \rangle$ via $\xi_i = (\partial_i \theta - a_i)\,|\mathcal{O}|/\max|\mathcal{O}|$, and the normal-fluid velocity $\mathbf{v}^n$ from the constitutive relations of $\langle T^{\mu\nu} \rangle$ and $\langle J^\mu \rangle$. Consequently, the energy spectra $E_s(k)$ and $E_n(k)$, alongside all two-fluid velocity statistics, are evaluated self-consistently (see Methods).

Unlike in three dimensions, maintaining statistically steady 2D quantum turbulence requires continuous, homogeneous vortex injection. We achieve this via the Landau instability~\cite{arean2024hydrodynamics}.
By imposing a global mean superfluid velocity $\bar{\xi}_x$ above the Landau critical threshold, we continuously drive vortex nucleation and annihilation throughout the entire simulation domain, without introducing any inhomogeneous external force that would otherwise contaminate the velocity statistics.

To confirm that the system relaxes into a genuine first-principles turbulent steady state, we monitor the spatially averaged physical observables over time. As shown in Figs.~\ref{fig:tevol_driven}(a) and~\ref{fig:tevol_driven}(b), the energy spectra \(E_s(k)\) and \(E_n(k)\) become statistically stationary after an initial transient. This is corroborated in Fig.~\ref{fig:tevol_driven}(c), where \(\bar{\xi}_x\), \(N_v\), \(\bar{T}\), and \(\bar{v}_x^n\) converge to plateau values with only small fluctuations. 
The simultaneous growth of $\bar v_x^n$ and $N_v$ provides direct evidence of mutual-friction-induced drag on the normal fluid.
The vortex number stabilizes through a sustained balance between nucleation and annihilation, while the temperature remains essentially constant. The condensate magnitude at the steady state is shown in Fig.~\ref{fig:tevol_driven}(d). 
In this work, all quantities are measured in units of the mean chemical potential $\bar\mu$, which we set to unity.

\begin{figure*}[htbp]
\centering
\includegraphics[width=1.01\textwidth]{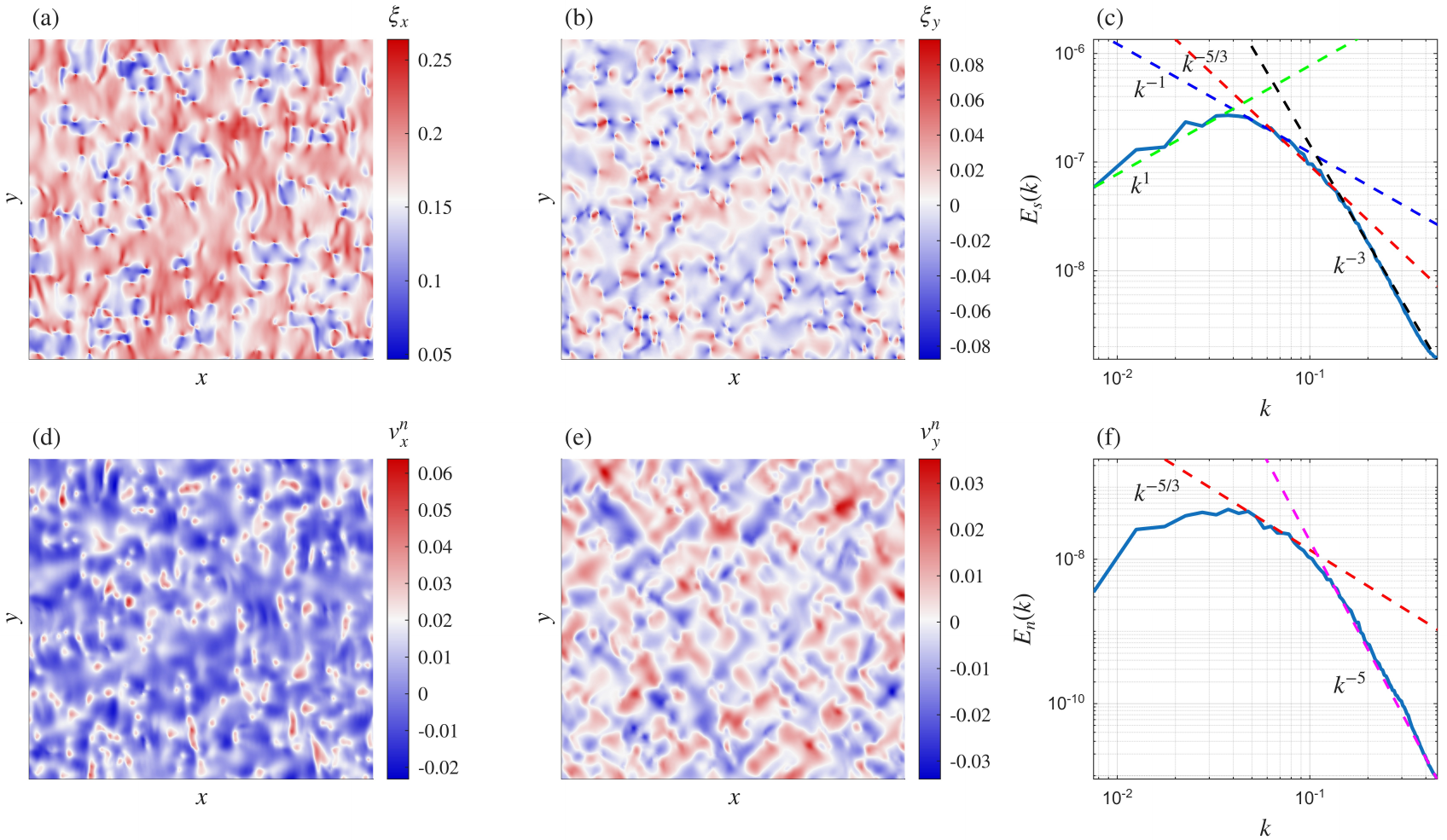}
\caption{\textbf{Steady-state velocity fields and energy spectra.} \textbf{(a,b)} Superfluid velocity components \(\xi_x\) and \(\xi_y\) at \(t \simeq 9000\). \textbf{(c)} Superfluid energy spectrum \(E_s(k)\), time-averaged over 100 slices within \(t \in [6000, 9000]\), displaying \(k^1\) scaling at large scales, \(k^{-1}\) and \(k^{-5/3}\) at intermediate scales, and \(k^{-3}\) at small scales. \textbf{(d,e)} Normal-fluid velocity components \(v_x^n\) and \(v_y^n\) at \(t \simeq 9000\). \textbf{(f)} Normal-fluid energy spectrum \(E_n(k)\), averaged over the same interval, showing \(k^{-5/3}\) scaling at intermediate scales and \(k^{-5}\) at small scales. Reference power laws are indicated by dashed lines. For both the superfluid and normal fluid, the spatial distributions of the longitudinal ($x$) and transverse ($y$) velocity components display markedly different structural features, revealing a clear spatial anisotropy. The system is in a statistically steady state with $\bar T/T_c=0.635$ and $\bar\xi_x=0.17$. 
\label{fig:v&E}}
\end{figure*}

\begin{figure*}[htbp]
\centering
\includegraphics[width=1.0\textwidth]{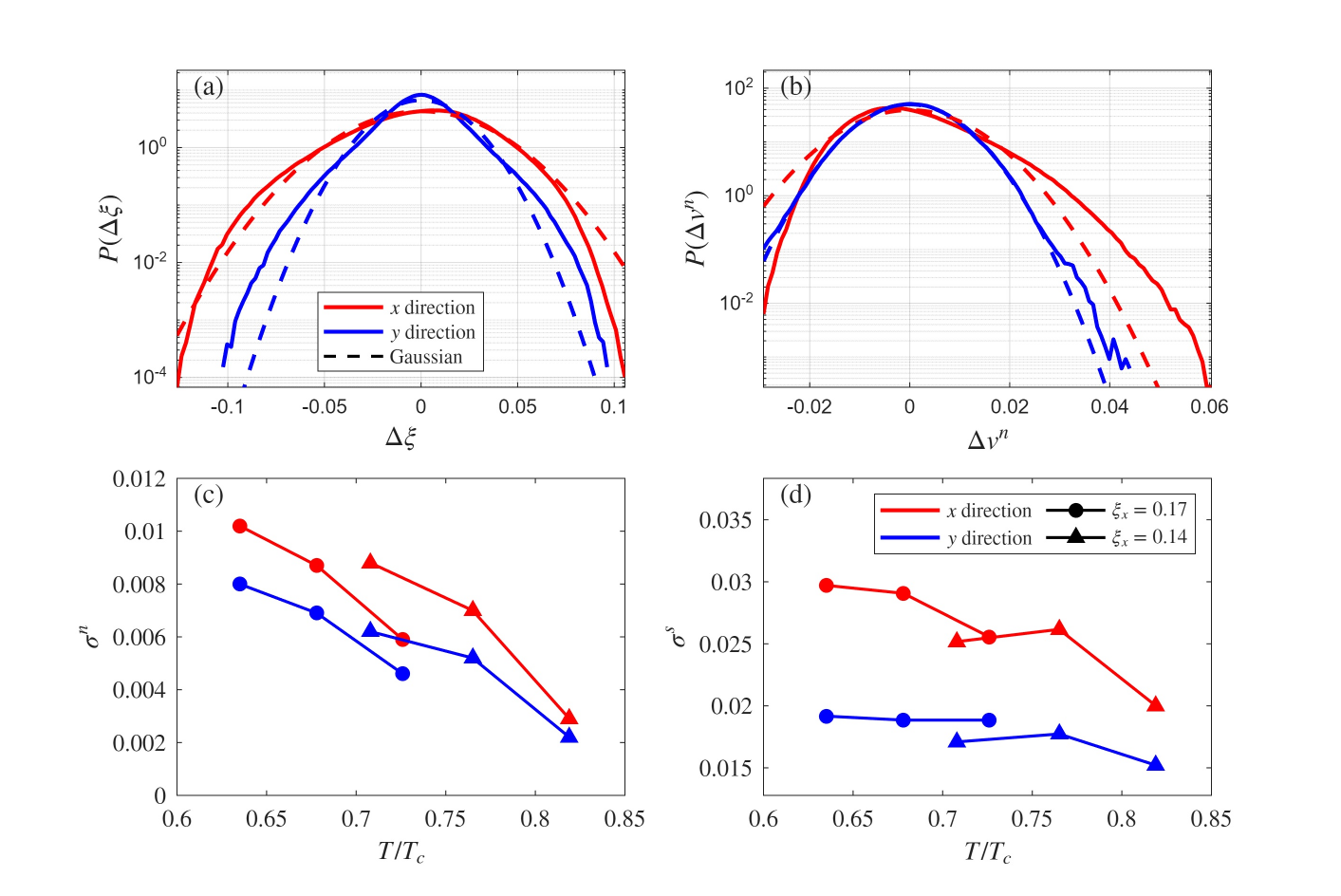}
\caption{\textbf{Velocity distributions and temperature dependence of fluctuations.} \textbf{(a, b)} Probability density functions (PDFs) of velocity fluctuations for the superfluid and normal fluid, \(P(\Delta \xi)\) and \(P(\Delta v^n)\) with $\Delta \xi=\xi-\bar\xi, \Delta v^n=v^n-\bar v^n$, accumulated over 100 time slices in the steady state. Both components exhibit pronounced non-Gaussian tails and spatial anisotropy, with fluctuations along the driving direction (\(x\)) significantly stronger than in the transverse direction (\(y\)); dashed lines denote Gaussian fits for reference. \textbf{(c, d)} Standard deviations of the normal-fluid and superfluid velocity fluctuations, \(\sigma^n\) and \(\sigma^s\), as functions of temperature and driving velocity \(\bar\xi_x\). The normal-fluid fluctuations \(\sigma^n\) decrease monotonically with increasing temperature across all parameter sets, despite the accompanying increase in vortex density. This robust suppression is a direct consequence of thermal depletion of the superfluid density, which weakens the dynamical backreaction of vortices onto the normal fluid. In contrast, the superfluid fluctuations \(\sigma^s\) exhibit a non-monotonic temperature dependence, reflecting the competing effects of increasing vortex density and decreasing superfluid density. \label{fig:driven}}
\end{figure*}

\subsection*{Spectral Signatures of Active Backreaction}

Having established a first-principles steady turbulent state, we now turn to its spectral and statistical characterisation. Representative snapshots of the superfluid and normal-fluid velocity fields at $t \simeq 9000$ are presented in Figs.~\ref{fig:v&E}(a,b) and (d,e), illustrating the complex coupling between the two components induced by mutual friction. In the statistically steady driven state, the superfluid energy spectrum \(E_s(k)\) displays a cascade of scaling regions characteristic of 2D quantum turbulence (Fig.~\ref{fig:v&E}(c)). 
At large scales, the spectrum scales as \(k^{1}\), indicating a random velocity distribution—a feature that broadens with increasing vortex density (Supplementary Figs.~S5 and S6). At intermediate scales, we observe a short \(k^{-1}\) region arising from the single-vortex velocity profile, followed by a Kolmogorov \(k^{-5/3}\) inertial range, whose extent is truncated by the proximity of neighbouring vortices. At small scales, the spectrum follows a \(k^{-3}\) scaling governed by the vortex-core structure. The normal-fluid energy spectrum \(E_n(k)\) (Fig.~\ref{fig:v&E}(f)) exhibits a short \(k^{-5/3}\) intermediate region and a \(k^{-5}\) scaling at small scales, consistent with pure holographic normal-fluid turbulence~\cite{adams2014holographic}. This behaviour gradually deviates at higher temperatures as vortex cores overlap and individual signatures become smeared out (Supplementary Figs.~S5 and S6).

While the energy spectra confirm the turbulent nature of the flow, the most striking signatures of the vortex–normal-fluid coupling emerge in the real-space velocity distributions. Figs.~\ref{fig:driven}(a,b) present the probability density functions (PDFs) of the velocity fluctuations for both components. Remarkably, the normal-fluid velocity PDFs exhibit pronounced non-Gaussian tails and a clear spatial anisotropy: fluctuations along the driving direction ($x$) are significantly enhanced compared to those in the transverse direction ($y$), aligning closely with recent experimental and numerical observations~\cite{mastracci2019particle,yui2020fully}. These non-Gaussian, anisotropic signatures stem directly from the dynamical backreaction of quantized vortices on the normal fluid. Uncoupled, a pure normal fluid exhibits a Gaussian velocity distribution. Under a global mean flow, however, vortices preferentially imprint their fluctuations along the superfluid drift, generating a statistical anisotropy fundamentally inaccessible to passive thermal-bath descriptions. The superfluid velocity PDFs likewise show non-Gaussian features, though with a different anisotropy structure reflecting the vortex-dominated kinematics. 
A $\Delta v^{-3}$ power-law tail, typically expected from single-vortex velocity statistics~\cite{paoletti2008velocity,white2010nonclassical,adachi2011numerical,baggaley2011quantum}, is absent in the driven case due to high vortex density, but clearly emerges during late-stage decaying turbulence when vortices are well separated (Supplementary Fig. S4).

Crucially, these non-Gaussianities persist across all parameter sets we have investigated, and their magnitude exhibits a systematic temperature dependence. As the temperature is raised, the vortex density increases to the point where the mean inter-vortex distance approaches the vortex core size itself (Supplementary Figs.~S5 and S6). This spatial crowding effectively smears out the sharp statistical signatures that would otherwise be induced by highly isolated individual vortex cores, leading to a gradual reduction in the non-Gaussianity of the normal-fluid velocity distributions. 

The anisotropic, non-Gaussian velocity statistics presented here establish that the normal fluid in a holographic superfluid is far from a passive thermal reservoir. Instead, it actively responds to the quantized vortex dynamics, acquiring a distinct statistical signature that directly reflects the backreaction of the superfluid component. This provides a first-principles confirmation, within a fully consistent theoretical framework, of the active normal-fluid behaviour recently observed in superfluid $^4$He experiments~\cite{marakov2015visualization,gao2017energy,mastracci2018exploration,mastracci2019particle}.

\subsection*{Temperature Suppression of Normal-Fluid Fluctuations}

Alongside the geometric smearing of the non-Gaussian
tails, the driven state exhibits a striking temperature-dependent suppression of normal-fluid velocity fluctuations \(\sigma_x^n\) and \(\sigma_y^n\) (Fig.~\ref{fig:driven}(c)) despite a simultaneous increase in vortex density. To determine whether this suppression is an genuine finite-temperature effect or merely a consequence of vortex crowding, we compare the driven regime with freely decaying turbulence, where the system is initialized above the Landau critical velocity and then relaxes through vortex--antivortex annihilation without continuous driving. 
In the decaying regime, vortex density drops while temperature remains constant, leading to a concurrent decrease in $\sigma^n$ (Supplementary Figs.~S2 and S3). Conversely, in the driven regime, $\sigma^n$ decreases with rising temperature even as vortex density increases. The opposite trends rule out vortex density as the dominant control parameter and identify an genuine finite-temperature suppression.

What is the physical mechanism underlying this suppression? The answer lies in the thermal depletion of the superfluid density $\rho_s$. At higher temperatures, the superfluid condensate is progressively depleted as the system approaches $T_c$. Since the dynamical backreaction exerted by quantized vortices on the normal fluid is mediated by the superfluid component—the vortices are topological defects in the superfluid order parameter, and their influence on the normal fluid scales with the strength of the condensate—a reduction in $\rho_s$ inherently weakens the coupling between the two fluids. Consequently, even though the vortex number increases, each individual vortex exerts a weaker drag on the normal fluid, leading to a net suppression of normal-fluid velocity fluctuations. This physical picture is further corroborated by the behavior of the superfluid velocity fluctuations, which, unlike those of the normal fluid, exhibit a non-monotonic dependence on temperature (Fig.~\ref{fig:driven}(d)), reflecting the competing effects of increasing vortex density and decreasing condensate stiffness.

The decoupling of temperature effects from vortex-density effects achieved here represents a key conceptual advance. In conventional hybrid approaches, the mutual friction coefficients are treated as temperature-dependent empirical parameters that are calibrated against equilibrium data, making it impossible to unambiguously separate thermal effects from the geometric effects of vortex crowding. Our fully backreacted holographic framework, by contrast, encodes both contributions from first principles, allowing us to isolate the genuine temperature-driven suppression of normal-fluid backreaction. This finding establishes that the active role of the normal fluid is not solely determined by the number of vortices present, but is fundamentally governed by the superfluid condensate itself.

\subsection*{Implications and Outlook}

Our results establish fully backreacted holographic superfluids as a unified, first-principles paradigm for quantum turbulence, completely circumventing the Lagrangian--Eulerian inconsistencies and phenomenological fitting parameters of conventional hybrid models. By intrinsically capturing momentum transfer, finite-temperature dissipation, and mutual friction from the gravitational dynamics, our framework provides an exact theoretical laboratory for exploring the far-from-equilibrium behaviour of strongly coupled quantum fluids. 
We establish that the normal fluid is an active participant driven by quantized vortex backreaction, yet its macroscopic fluctuations are fundamentally governed by the thermal depletion of the superfluid condensate rather than simple vortex crowding.

A key advantage of the holographic description is that the full nonlinear gravitational dynamics automatically incorporates all orders of derivative corrections, without the need to introduce explicit transport coefficients.
This stands in sharp contrast to the standard two-way coupled simulations, where the mutual friction coefficients must be either calibrated against equilibrium experiments or estimated from theoretical models of vortex--excitation scattering. 
And in contrast to probe limit, our fully backreacted framework
dynamically generates the normal-fluid velocity field, enabling the first ab initio extraction of its non-Gaussian PDFs and their temperature dependence. The qualitative agreement of these signatures with observations in superfluid~\cite{marakov2015visualization,gao2017energy,mastracci2018exploration,mastracci2019particle} without any fitted parameters provides strong evidence that the holographic approach captures the essential physics of quantum turbulence at a fundamental level.

The immediate applicability of our 2D setup extends naturally to ongoing experiments on ultra-cold atomic superfluids and superfluid films confined to planar geometries. The quantitative predictions emerging from our velocity statistics—for instance, the kurtosis of the normal-fluid PDF as a function of the superfluid fraction \(\rho_s/\rho\)—provide testable benchmarks for particle-tracking velocimetry and particle-image velocimetry experiments in these systems. Furthermore, our steady-state setup opens the door to investigating how active two-fluid coupling affects the disordered hyperuniformity recently identified in 2D steady vortex matter~\cite{An:2026bcw}. Looking ahead, the extension of this fully backreacted framework to three spatial dimensions represents the most consequential and immediate frontier, which is in principle straightforward by promoting the spatial coordinates from \((x,y)\) to \((x,y,z)\) and allowing for more general metric ansätze.
Such an extension would enable, for the first time, a first-principles investigation of vortex-line reconnections, the precise form of the mutual friction force in the fully nonlinear regime, and the detailed scaling of the energy spectrum across the entire inertial range (see~\cite{wittmer2024quantum,zeng2025dissipation} for previous study in probe limit). Ultimately, this would provide a unified, exact theoretical perspective on quantum turbulence that spans from the microscale vortex-core dynamics to the macroscale hydrodynamic transport, applicable to systems as diverse as superfluid helium, cold atomic gases, and neutron-star interiors.

More broadly, our work demonstrates that holographic duality is a powerful dynamical framework for strongly coupled quantum matter far from equilibrium.
The methodological advances reported here—including the fully backreacted numerical evolution with dynamical horizon tracking and the self-consistent extraction of two-fluid observables—lay the groundwork for a wide range of future applications beyond quantum turbulence including the real-time dynamics of inhomogeneous condensates, the interplay between superfluidity and magnetic fields in holographic superconductors, and the non-equilibrium response of strongly coupled phases to external driving. By bridging the gap between gravitational simulations and experimentally accessible quantum fluid phenomena, the holographic approach opens a new avenue for understanding the emergent collective behaviour of strongly interacting quantum matter in its most complex and unpredictable regimes.

\section*{Methods}

\subsection*{Holographic setup and equations of motion}

Varying the action~\eqref{eq:action} yields the coupled Einstein, Maxwell and scalar equations of motion:
\begin{align}
&R_{ab}-\frac{1}{2}g_{ab}R-\frac{3}{L^2}g_{ab}=8\pi G\, T_{ab}\,,\\
&\nabla_a F^{ab}=iq\left[
\Psi^* D^b \Psi-
\Psi (D^b\Psi)^*\right]\,,\\
&D_a D^a \Psi - m^2 \Psi = 0\,.
\end{align}
with the energy-momentum tensor
\begin{equation}
    \begin{aligned}
    T_{ab}=&\frac{1}{2}[F_{ac}F_b{}^c-\frac{1}{4}g_{ab}F^2+(D_a\Psi)^* D_b\Psi\\&+(D_b\Psi)^* D_a\Psi-g_{ab}\left(|D\Psi|^2 + m^2 |\Psi|^2\right)]\,.
    \end{aligned}
\end{equation}

To simulate real-time dynamics without symmetry reduction, we adopt the ingoing Eddington-Finkelstein metric ansatz:
\begin{align}
ds^2=&-2Adt^2-\frac{2}{z^2}dtdz-2F_xdtdx-2F_ydtdy\nonumber\\&+\Sigma^2\Big[e^{B}\cosh C\, dx^2+2\sinh C\, dxdy\nonumber\\&+e^{-B}\cosh C\, dy^2\Big]\,,\\
A_\mu dx^\mu&=A_t dt+A_xdx+A_y dy,\quad \Psi=\Psi_{R}+i\Psi_{I}\,,\nonumber
\label{eq:metric_ansatz}
\end{align}
where the radial gauge \(A_z=0\) is chosen, and \(\Psi_R\) and \(\Psi_I\) denote the real and imaginary parts of \(\Psi\). There are 11 real functions of \((t,z,x,y)\). Altogether, these 11 real functions satisfy 16 coupled, nonlinear PDEs in the same coordinates. This ansatz captures fully nonlinear and anisotropic dynamics in two spatial dimensions. The apparent horizon is fixed at a constant $z_h$ throughout the evolution by a suitable choice of the residual diffeomorphism gauge (see Supplementary Information for details).

Since we consider fully nonlinear time evolution, the bulk geometry corresponds to a dynamical black hole spacetime rather than a static equilibrium configuration. Although a dynamical spacetime does not possess a globally well-defined equilibrium temperature, one can still define an effective local temperature from the near-horizon geometry 
\begin{equation}\label{eq:T}
T(t,x,y)=-\frac{z_h^2}{2\pi }\partial_zA|_{z_h}\,.
\end{equation}
In equilibrium, this reduces to the standard Hawking temperature determined by the surface gravity of the stationary black hole. During time evolution, the horizon area changes dynamically and encodes entropy production in the dual strongly coupled field theory.

Near the AdS boundary $z \to 0$, the fields admit asymptotic expansions. The scalar field $\Psi = \Psi_R + i\Psi_I$ expands as
\begin{equation}
\Psi_R = z\Psi_{R1} + z^2\Psi_{R2} + \cdots\,,\,\,
\Psi_I = z\Psi_{I1} + z^2\Psi_{I2} + \cdots\,,
\end{equation}
and the gauge field as $A_\mu = a_\mu + z a_{1\mu} + \cdots$ in radial gauge $A_z = 0$. Imposing source-free conditions $\Psi_{R1} = \Psi_{I1} = 0$ realises spontaneous $U(1)$ breaking. The superfluid order parameter is then $\langle \mathcal{O}\rangle = \Psi_{R2} + i\Psi_{I2}$, the conserved current is $\langle J^\mu\rangle = a_1^\mu$, and the stress-energy tensor $\langle T^{\mu\nu}\rangle$ is obtained from the subleading metric coefficients via holographic renormalization~\cite{arean2021holographic,arean2024hydrodynamics}. 

In this work, we set $L=1$, $G=1/16\pi$, $q=1$ and $m^2=-2$ without loss of generality, and only consider standard quantization.

\subsection*{Extraction of hydrodynamic observables}

The superfluid velocity is extracted from the order parameter as
\begin{equation}
\xi_\mu = (\partial_\mu\theta - a_\mu)\frac{|\mathcal{O}|}{\max|\mathcal{O}|}\,,
\end{equation}
where the last factor regularises the singularity at vortex cores. The normal-fluid velocity $u^\mu$ is obtained from the ideal-fluid constitutive relations~\cite{sonner2010gravity,bhattacharya2011dissipative}
\begin{equation}
T^{\mu\nu} = (\epsilon + p)u^\mu u^\nu + p\eta^{\mu\nu} + \alpha \xi^\mu\xi^\nu,\;
J^\mu = (\rho - \rho_s)u^\mu + \alpha \xi^\mu\,,
\end{equation}
with $u^\mu u_\mu = -1$, $u^\mu\xi_\mu + \mu = 0$ and $\alpha=\rho_s/\mu$. Here $\epsilon$ is the energy density, $p$ is the pressure, $\rho$ is the total charge density (number density), and $\rho_s$ is the superfluid denisity. 

We define \(Q^\mu \equiv J^\mu - \alpha\xi^\mu\), which yields
\begin{equation}\label{eq:u_reconstruction}
u^\mu = \frac{Q^\mu}{\sqrt{-Q^2}}\,.
\end{equation}
The energy density \(\epsilon\) and parameter \(\alpha\) are then determined from the stress tensor, and the superfluid density follows as \(\rho_s = \alpha\mu\). This ideal-fluid extraction is a leading-order approximation in derivatives. In the presence of vortices, derivative corrections are not expected to yield a definitively more reliable description, and we therefore restrict ourselves to~\eqref{eq:u_reconstruction} for the normal-fluid velocity throughout this work. Further details, including the treatment of derivative corrections, are provided in Supplementary Information.

\subsection*{Driving protocol for sustained turbulence}

To sustain a statistically steady turbulent state in two dimensions, we exploit the Landau instability~\cite{arean2024hydrodynamics} by imposing a global mean superfluid velocity \(\bar{\xi}_x\) above the critical threshold. In practice, we fix the spatial component \(a_x\) of the gauge field and initialise the phase gradient as \(\partial_x\theta(t=0)=0\). At each time step, we monitor the winding number along the \(x\)-direction:
\begin{equation}
W = \mathrm{round}\left(\frac{1}{2\pi L_y}\int_0^{L_y}dy\int_0^{L_x}\partial_x\theta\,dx\right),
\end{equation}
where the integral is averaged over the transverse \(y\)-direction and constrained to be an integer by the periodic boundary conditions. Whenever \(|W| \ge 1\), we apply a global phase shift
\begin{equation}
\Psi \to \Psi e^{-i2\pi W x/L_x},
\end{equation}
which resets the winding number while preserving the mean superfluid velocity \(\bar{\xi}_x = -a_x\). This procedure prevents the turbulent state from decaying via vortex annihilation and does not alter the energy spectra or velocity fluctuations.
Crucially, these instantaneous changes to the phase of the scalar field—a propagating degree of freedom—do not violate the bulk Einstein equations. The phase resets merely amount to introducing these targeted boundary conditions to the bulk equations: 
\begin{align}
\partial_t A_x(z=0) = \mathrm{const.} \times \sum_i \delta(t-t_i)\,,\\
\partial_x A_t(z=0) = \mathrm{const.} \times \sum_i \delta(t-t_i)\,,
\end{align}
 where $t_i$ denotes the specific times at which the phase is shifted.
Physically, this scheme is equivalent to driving the system via delta-function pulses in the chemical potential, $\mu = x\,\mathrm{const.} \times \sum_i \delta(t-t_i)$, following from the Josephson relation.

\subsection*{Numerical implementation}

The equations of motion are discretised using a pseudo-spectral method. Fourier grids with \(200\) points in each spatial direction and a Chebyshev grid with \(20\) points in the radial direction are used. Time evolution employs fourth-order Runge--Kutta with step size $dt = 0.05$ in units of $z_h$. The simulation domain is $L_x\times L_y=400 \times 400$ in units of $z_h$. Further details of the time evolution scheme, including the nested radial integration and the constraint equations, are provided in Supplementary Information.



\section*{Acknowledgements}
We thank Amos Yarom for the initial collaboration and valuable discussions.
YA was supported by the Israel Science Foundation (ISF) Grants No.\,3191/23 and No.\,2916/24 and by the United States-Israel Binational Science Foundation (BSF) Grant No.\,2034142. LL was supported by the National Natural Science Foundation of China Grants No.\,12525503, No. 12588101, and No.\,12447101. MT was supported by JSPS KAKENHI Grant No. JP22H05139 and No. JP22H05131. HZ was supported by the National Natural Science Foundation of China Grants No.\,11675140.
The work of SW was supported by the Kreitman School of Advanced Graduate Studies through a Kreitman Fellowship, by the Israel Science Foundation (grant No. 1417/21), by the German Research Foundation through a German-Israeli Project Cooperation (DIP) grant “Holography and the Swampland”, by Carole and Marcus Weinstein through the BGU Presidential Faculty Recruitment Fund, by the ISF Center of Excellence for theoretical high energy physics, by the VATAT Research Hub in the Field of quantum computing and by the ERC starting Grant dSHologQI (project number 101117338).
\bibliography{biblio.bib}



\clearpage
\onecolumngrid 

\setcounter{equation}{0}
\setcounter{figure}{0}
\setcounter{table}{0}
\setcounter{page}{1}
\setcounter{section}{0}

\renewcommand{\theequation}{S\arabic{equation}}
\renewcommand{\thefigure}{S\arabic{figure}}
\renewcommand{\thetable}{S\arabic{table}}

\begin{center}
\textbf{\Large Supplementary Information for: \\ Active normal-fluid feedback in quantum turbulence from holography}
\end{center}
\vspace{2em}

\section{Detailed Time Evolution Scheme and Nested Equations}\label{app:numerical details}

We use a characteristic formalism similar to~\cite{chesler2014numerical} to study time evolution of fully backreacted holographic superfluids.
Following Chesler-Yaffe's notation, we define
\begin{equation}
    d_+f=\partial_tf-z^2A\partial_zf\,,
\end{equation}
and denote the equations of motion (EoMs) as
\begin{align}
&E_{ab}\equiv R_{ab}
-
\frac{1}{2}g_{ab}R
-
\frac{3}{L^2}g_{ab}
-
8\pi G\, T_{ab}=0\,,\\
&V^{b}\equiv\nabla_a F^{ab}
-
iq
\left[
\Psi^* D^b \Psi
-
\Psi (D^b\Psi)^*
\right]=0\,,\\
&O\equiv D_a D^a \Psi - m^2 \Psi = 0\,.
\end{align}
Below we set $G=1/16\pi$. By taking certain combinations, we can write EoMs in nested form as a set of ordinary differential equations that can be solved in order. To be specific:
\begin{equation}\label{eq:Sigma}
\begin{aligned}
    &\Sigma: E_{zz}=(\partial_z^2+\frac{2}{z}\partial_z+Q_\Sigma[f_\Sigma])\Sigma=0\,,\\&\qquad f_\Sigma=\{B,C,A_x,A_y,\Psi\}\,,
    \end{aligned}
\end{equation}
\begin{equation}\label{eq:F}
\begin{aligned}
    &F_i: E_{zi}+\frac{z^2}{2}g_{ti}E_{zz}=(\delta_i^j\partial_z^2+P_{F}[f_F]_i^j\partial_z+Q_{F}[f_F]_i^j)F_j-S_F[f_F]_i=0\,,\\&\qquad f_F=\{B,C,A_x,A_y,\Psi,\Sigma\}\,,
    \end{aligned}
\end{equation}
\begin{equation}\label{eq:At}
\begin{aligned}
    &A_t: V^{z}=(\partial_z^2+P_{A_t}[f_{A_t}]\partial_z)A_t-S_{A_t}[f_{A_t}]=0\,,\\&\qquad f_{A_t}=\{B,C,A_x,A_y,\Psi,\Sigma,F_i\}\,,
    \end{aligned}
\end{equation}
\begin{equation}\label{eq:dSigma}
\begin{aligned}
    &d_+\Sigma: E_{tz}+\frac{z^2}{2}g_{tt}E_{zz}=(\partial_z+Q_{d_+\Sigma}[f_{d_+\Sigma}])d_+\Sigma-S_{d_+\Sigma}[f_{d_+\Sigma}]=0\,,\\&\qquad f_{d_+\Sigma}=\{B,C,A_x,A_y,\Psi,\Sigma,F_i,A_t\}\,,
    \end{aligned}
\end{equation}
\begin{equation}\label{eq:dPhi}
\begin{aligned}
    &\{d_+\Psi_R,\,d_+\Psi_I\}: \{O+O^*,-i(O-O^*)\}=(\partial_z+Q_{d_+\Psi}[f_{d_+\Psi}])d_+\Psi-S_{d_+\Psi}[f_{d_+\Psi}]=0\,,\\&\qquad f_{d_+\Psi}=\{B,C,A_x,A_y,\Psi,\Sigma,F_i,A_t,d_+\Sigma\}\,,
    \end{aligned}
\end{equation}
\begin{equation}\label{eq:dBdCCdAxdAy}
\begin{aligned}
    &\{d_+B,\,d_+C,\,d_+A_x,\,d_+A_y\}: \{E_{yy}-e^{-2B}E_{xx},\, E_{xy}-e^{-B}\tanh\,C\,E_{xx},\,V^x,\,V^y\}=\\&\qquad(P_{d_+f}[f_{d_+f}]^I_J\partial_z+Q_{d_+f}[f_{d_+f}]^I_J)d_+f^J-S_{d_+f}[f_{d_+f}]^I=0\,,\\&\qquad d_+f_I=\{d_+B,\,d_+C,\,d_+A_x,\,d_+A_y\}\,,\\&\qquad f_{d_+f}=\{B,C,A_x,A_y,\Psi,\Sigma,F_i,A_t,d_+\Sigma,d_+\Psi\}\,,
    \end{aligned}
\end{equation}
\begin{equation}\label{eq:A}
\begin{aligned}
    &A: E_{tz}-\frac{z^2}{2}g_{tt}E_{zz}=(\partial_z^2+P_{A}[f_{A}]\partial_z+Q_A[f_A])A-S_{A}[f_{A}]=0\,,\\&\qquad f_{A}=\{B,C,A_x,A_y,\Psi,\Sigma,F_i,A_t,d_+\Sigma,d_+\Psi,d_+B,\,d_+C,\,d_+A_x,\,d_+A_y\}\,,
    \end{aligned}
\end{equation}
\begin{equation}\label{eq:dF}
\begin{aligned}
    &d_+F_i: E_{ti}+\frac{z^2}{2}g_{tt}E_{zi}+z^2g_{ti}(E_{tz}+\frac{z^2}{2}g_{tt}E_{zz})=\\&\qquad(\delta_i^j\partial_z+Q_{d_+F}[f_{d_+F}]_i^j)d_+F_j-S_{d_+F}[f_{d_+F}]_i=0\,,\\&\qquad f_{d_+F}=\{B,C,A_x,A_y,\Psi,\Sigma,F_i,A_t,d_+\Sigma,d_+\Psi,d_+B,\,d_+C,\,d_+A_x,\,d_+A_y,A\}\,,\\
\end{aligned}
\end{equation}
\begin{equation}\label{eq:ddSigma}
\begin{aligned}
    &d_+d_+\Sigma: E_{tz}+\frac{z^2}{2}g_{tt}E_{zz}=d_+d_+\Sigma-S_{d_+d_+\Sigma}[f_{d_+d_+\Sigma}]=0\,,\\&\qquad f_{d_+d_+\Sigma}=\{B,C,A_x,A_y,\Psi,\Sigma,F_i,A_t,d_+\Sigma,d_+\Psi,d_+B,\,d_+C,\,d_+A_x,\,d_+A_y,A\}\,,
\end{aligned}
\end{equation}
\begin{equation}\label{eq:dAt}
\begin{aligned}
    &dA_t: V^{t}=\partial_zd_+A_t-S_{dA_t}[f_{dA_t}]=0\,,\\&\qquad f_{dA_t}=\{B,C,A_x,A_y,\Psi,\Sigma,F_i,A_t,d_+\Sigma,d_+\Psi,\,d_+A_x,\,d_+A_y,A,dF_i\}\,,
    \end{aligned}
\end{equation}
The explicit forms of $P$, $Q$ and $S$ are extremely lengthy and not instructive, so we omit them. But the calculation is straight forward.

Firstly, we need $\{B,C,A_x,A_y,\Psi\}$ as initial conditions. There's no constraint for initial profiles of these fields. With these fields at hand, we can solve the nested EoMs to get $\{\Sigma,F_i,A_t,d_+\Sigma,d_+\Psi,d_+B,\,d_+C,\,d_+A_x,\,d_+A_y,A\}$. Time derivatives of fields are given by 
\begin{equation}
    \partial_t f=d_+f+z^2A\partial_zf\,.
\end{equation}
Now we can evolve $\{B,C,A_x,A_y,\Psi\}$ to next time step and then repeat this procedure. In practice we only need to solve~\eqref{eq:dF}, \eqref{eq:ddSigma} and~\eqref{eq:dAt}
on the AdS boundary and they would be automatically satisfied in the bulk ensured by other equations. On AdS boundary they give momentum, energy and charge conservation equations of the boundary field theory respectively:
\begin{align}
\label{eq:a3t}
    &\partial_ta_3=\frac{3}{4}\partial_if_{i3}-\frac{1}{4}[(\partial_xa_t-\partial_ta_x)(\partial_xa_t-\partial_ta_x+a_{1x})+(\partial_ya_t-\partial_ta_y)(\partial_ya_t-\partial_ta_y+a_{1y})]\,,\\
    \label{eq:fx3t}
    &\partial_t f_{x3}=\frac{1}{3}[\partial_x(2a_3-3b_3)-3\partial_yC_3+(a_{1t}(\partial_ta_{x}-\partial_xa_{t})-(\partial_y a_{t}-\partial_t a_{y}+a_{1y})(\partial_ya_x-\partial_xa_y)]\,,\\
    \label{eq:fy3t}
    &\partial_t f_{y3}=\frac{1}{3}[\partial_x(2a_3+3b_3)-3\partial_xC_3+(a_{1t}(\partial_ta_{y}-\partial_ya_{t})+(\partial_x a_{t}-\partial_t a_{x}+a_{1x})(\partial_ya_x-\partial_xa_y)]\,,\\
    \label{eq:a1t}
    &\partial_ta_{1t}=\partial_i(a_{1i}+\partial_ia_t-\partial_ta_i)\,.
\end{align}
If we turn on scalar source there would be more terms.

Near AdS boundary, the metric functions are expanded as
\begin{equation}
A=\frac{1}{2z^2}+\frac{\Sigma_0(t,x,y)}{z}+\frac{1}{2}\Sigma_0^2-\partial_t\Sigma_0+za_3(t,x,y)+\cdots,
\end{equation}
\begin{equation}
F_x=-\partial_x\Sigma_0(t,x,y)+zf_{x3}(t,x,y)+\cdots,
\end{equation}
\begin{equation}
F_y=-\partial_y\Sigma_0(t,x,y)+zf_{y3}(t,x,y)+\cdots,
\end{equation}
\begin{equation}
\Sigma=\frac{1}{z}+\Sigma_0(t,x,y)+z^3\Sigma_3(t,x,y)+\cdots,
\end{equation}
\begin{equation}
B=z^3b_3(t,x,y)+\cdots,
\qquad
C=z^3C_3(t,x,y)+\cdots.
\end{equation}
The coefficients appearing in the asymptotic expansion of the metric determine the expectation value of the boundary stress-energy tensor through holographic renormalization. In particular, \(a_3\), \((f_{x3},\,f_{y3})\), \(b_3\) and \(C_3\) encode the local energy density, momentum density, pressure, and anisotropic stress of the dual field theory. 

The function \(\Sigma_0(t,x,y)\) originates from the residual radial diffeomorphism symmetry
\begin{equation}
z\rightarrow \frac{z}{1+z\Sigma_0(t,x,y)},
\end{equation}
and corresponds to a gauge degree of freedom rather than a physical observable. In numerical evolutions, this freedom can be used to fix the apparent horizon at a constant radial position throughout the evolution. In practice it's convenient to choose $\Sigma_0$ so that the apparent horizon always located at the numerical boundary $z_h(t,x,y)= z_{max}=const$. To achieve this, firstly we need to solve the position of the apparent horizon for an general initial configuration at $t_0$. The apparent horizon is the position where the expansion $\theta_l$ of the outgoing null ray $l$ equals zero:
\begin{align}
    &\theta_l=h^{ab}\nabla_al_b=\frac{1}{2}h^{ab}\mathcal{L}_lh_{ab}=0\,,\\
    h_{ab}=g_{ab}&+l_an_b+l_bn_a\,,\quad l^2=n^2=0\,,  \quad l\cdot n=-1\,,
\end{align}
with $n^a\sim (\partial_z)^a$ being the ingoing null congruence of geodesics. Normalization of $l$ and $n$ is not unique. We choose the normalization of $n$ to be $n^a=-(\partial_z)^a$. Together with $l_a(\partial_i)^a=0$, we can get
\begin{equation}
    l_a=(dz)_a+z^2(A+\frac{1}{2}F_iF_jh^{ij})(dt)_a\,.
\end{equation}
In the end the expansion is find to be
\begin{equation}\label{eq:expansion}
    \theta_l=\frac{2z^2}{\Sigma}\left[\partial_t+h^{ij}F_i\partial_j-z^2\left(A+\frac{1}{2}F_iF_jh^{ij}\right)\partial_z\right]\Sigma .
\end{equation}
For a stationary homogeneous black brane, $\theta_l=0$ is equivalent to the condition of Killing horizon. By requiring $\theta_l|_{z_{max}}=0$, we get an differential equation that can be used to solve $\Sigma_0(t_0,x,y)$. To fix the apparent horizon $z_h(t,x,y)=z_{max}$ for all moments, we also need to require $\partial_t\theta_l|_{z_{max}}=0$. After eliminating $\partial_t^2\Sigma$ and $\partial_t F_i$ by using the constraint equations~\eqref{eq:dF} and~\eqref{eq:ddSigma}, it turns out to be an elliptic equation of $A$ at horizon $z_h$. Once we get $A|_{z_h}$, $\partial_t\Sigma_0$ can be extracted from~\eqref{eq:redefinition_A} as
\begin{equation}
   \partial_t\Sigma_0 =(-A+\frac{1}{2z^2}+\frac{\Sigma_0}{z}+\frac{1}{2}\Sigma_0^2+\widetilde A)|_{z_h}\,.
\end{equation}
In our implementation, we actually solve $(\partial_t\theta_l+K\theta_l)|_{z_{max}}=0$ to improve numerical stability, following~\cite{bea2022holographic}, with $K$ a constant parameter.

To make the near-boundary behavior manifest and improve numerical stability, we separate the leading singular terms from the regular dynamical fields and make the following field redefinitions:
\begin{equation}\label{eq:redefinition_A}
A=\frac{1}{2z^2}+\frac{\Sigma_0(t,x,y)}{z}+\frac{1}{2}\Sigma_0^2-\partial_t\Sigma_0+\widetilde A(t,z,x,y),
\end{equation}
\begin{equation}
F_x=-\partial_x\Sigma_0(t,x,y)+\widetilde F_x(t,z,x,y),
\end{equation}
\begin{equation}
F_y=-\partial_y\Sigma_0(t,x,y)+\widetilde F_y(t,z,x,y),
\end{equation}
\begin{equation}
\Sigma=\frac{1}{z}+\Sigma_0(t,x,y)+z\widetilde\Sigma(t,z,x,y),
\end{equation}
\begin{equation}
B=z^2\widetilde B(t,z,x,y),
\qquad
C=z^2\widetilde C(t,z,x,y),
\end{equation}
\begin{equation}
\Psi_R=z\widetilde\Psi_R(t,z,x,y),
\qquad
\Psi_I=z\widetilde\Psi_I(t,z,x,y),
\end{equation}
\begin{equation}
A_t=\widetilde A_t(t,z,x,y),
\qquad
A_x=\widetilde A_x(t,z,x,y),
\qquad
A_y=\widetilde A_y(t,z,x,y),
\end{equation}
together with the boundary conditions at AdS boundary $z=0$:
\begin{align}
    &\widetilde A|_{z=0}=0,\quad\partial_z\widetilde A|_{z=0}=a_3,\\
    &\widetilde F_x|_{z=0}=0,\quad\partial_z\widetilde F_x|_{z=0}=f_{x3},\\
    &\widetilde F_y|_{z=0}=0,\quad\partial_z\widetilde F_y|_{z=0}=f_{y3},\\
    &\widetilde \Sigma|_{z=0}=0,\quad\partial_z\widetilde \Sigma|_{z=0}=0,\\
    &\widetilde B|_{z=0}=0,\quad\widetilde C|_{z=0}=0,\\
    &\widetilde \Psi_R|_{z=0}=0,\quad\partial_z\widetilde \Psi_I|_{z=0}=0,\\
    &\widetilde A_t|_{z=0}=a_t,\quad \partial_z\widetilde A_t|_{z=0}=a_{1t},\\
    &\widetilde A_x|_{z=0}=a_x,\quad \widetilde A_y|_{z=0}=a_y.
\end{align}

The equations of motion furthermore possess scaling symmetries which imply that not all parameters are physically independent. In particular, the system is invariant under the coordinate rescaling
\begin{equation}
(t,x,y,z)\rightarrow \lambda_0 (t,x,y,z),
\end{equation}
together with the gauge field transformation
\begin{equation}
(A_t,A_x,A_y)\rightarrow \frac{1}{\lambda_0}(A_t,A_x,A_y).
\end{equation}
Under this scaling transformation, physical quantities transform according to their scaling dimensions,
\begin{equation}
T\rightarrow \frac{T}{\lambda_0},
\qquad
\mu\rightarrow \frac{\mu}{\lambda_0},
\qquad
\rho\rightarrow \frac{\rho}{\lambda_0^2},
\qquad
\langle\mathcal O\rangle\rightarrow \frac{\langle\mathcal O\rangle}{\lambda_0^\Delta}.
\end{equation}
where $\Delta$ is the conformal dimension of the scalar operator. As a consequence, only dimensionless combinations are physically meaningful. One may therefore use the scaling symmetry to fix one overall scale in the problem, such as the horizon position or chemical potential. In this work, we set $z_h=1$ in simulations for convenience, and rescale all quantities in units of the mean chemical potential $\bar\mu$ such as
\begin{equation}
\frac{T}{\bar\mu},
\qquad
\frac{\langle\mathcal O\rangle}{\bar\mu^\Delta},
\qquad
\frac{\xi}{\bar\mu}.
\end{equation}
Solutions related by the scaling symmetry correspond to the same physical state and differ only by an overall choice of units.

\section{Holographic Renormalization}

To explicitly map the raw gravitational data to the macroscopic physical observables of the two-fluid model, we must first compute the exact expectation values of the boundary stress-energy tensor \(\langle T^{\mu\nu}\rangle\), the conserved current \(\langle J^\mu\rangle\), and the scalar condensate \(\langle \mathcal{O}\rangle\) by introducing appropriate boundary counterterms. 
To regulate infinities at AdS boundary, we add the usual counter terms to action~\eqref{eq:action} ~\cite{arean2021holographic,arean2024hydrodynamics}
\begin{equation}
S_{ct} =
\int d^3x \sqrt{-\gamma}
\left(
2K -4-R^{(\gamma)}-|\Psi|^2\right)\,,
\label{eq:counter_term}
\end{equation}
where $\gamma_{\mu\nu}$ is the induced metric at AdS boundary, $K_{\mu\nu}=\gamma_\mu^a\gamma_\nu^b\nabla_a n_b$ is the extrinsic curvature of AdS boundary with $n^\mu=-z\delta^\mu_z$ the normal vector, $K=\gamma^{\mu\nu}K_{\mu\nu}$ is its trace, and $R^{(\gamma)}$ is the Ricci scalar of $\gamma_{\mu\nu}$. So the total renormalized action is given by
\begin{equation}
    S_{ren}=S+S_{ct}\,.
\end{equation}
Variation of the renormalized on-shell action is given by
\begin{equation} \label{eq:variation}
    \begin{aligned}
        \delta S_{ren}^{os}=&\int d^3x \sqrt{-\gamma}
\bigg\{\delta\gamma_{\mu\nu}[ K^{\mu\nu}-(K+2+\frac{R^{(\gamma)}}{2}+\frac{|\Psi|^2}{2})\gamma^{\mu\nu} \\ & -(R^{(\gamma)})^{\mu\nu}] - n_\mu F^{\mu\nu}\delta A_\nu  -\Psi\delta\Psi^*-\Psi^*\delta\Psi \\& -g^{\mu\nu}n_\mu\left[\delta\Psi (\partial_\nu+iqA_\nu)\Psi^*+\delta\Psi^*(\partial_\nu-iqA_\nu)\Psi\right] \bigg\} |_{z=0}\,.
    \end{aligned}
\end{equation}
From this expression we find the expectation values of energy momentum tensor, charge current and scalar operator of the dual boundary field theory to be
\begin{equation}\label{eq:Tmunu}
    \begin{aligned}
            \langle T^{\mu\nu}\rangle&=\frac{2}{\sqrt{-\gamma^{(0)}}}\frac{\delta S^{os}_{ren}}{\delta\gamma_{\mu\nu}^{(0)}}\\&= \frac{2}{z^5}\big[K^{\mu\nu}-(K+2+\frac{R^{(\gamma)}}{2}+\frac{|\Psi|^2}{2})\gamma^{\mu\nu}-(R^{(\gamma)})^{\mu\nu}\big]|_{z=0}\,,
    \end{aligned}
\end{equation}
\begin{equation}
    \label{eq:Jmu}
    \langle J^\mu\rangle =\frac{1}{\sqrt{-\gamma^{(0)}}}\frac{\delta S^{os}_{ren}}{\delta A_{\mu}^{(0)}}= \frac{1}{z^3}\big[n_\nu F^{\nu\mu}\big]|_{z=0}\,,
\end{equation}
\begin{equation}
    \label{eq:O}
    \langle \mathcal{O}\rangle =\frac{1}{\sqrt{-\gamma^{(0)}}}\frac{\delta S^{os}_{ren}}{\delta \Psi^{(0)*}}=\frac{1}{z^2} \big[-\Psi-g^{\mu\nu}n_\mu(\partial_\nu-iqA_\nu)\Psi\big]|_{z=0}\,,
\end{equation}
with $\gamma^{(0)}_{\mu\nu}=\lim\limits_{z\rightarrow 0}z^2\gamma_{\mu\nu}$, $A^{(0)}_\mu=\lim\limits_{z\rightarrow 0}A_\mu$ and $\Psi^{(0)}=\lim\limits_{z\rightarrow 0}\Psi/z$. 
In terms of boundary expansion coefficients, these expectation values are
\begin{equation}\label{eq:Tmunu_expansion}
    \langle T^{\mu\nu}\rangle=3\left[
    \begin{array}{ccc}
       -\frac{4}{3}a_3  & f_{x3}  & f_{y3} \\
        f_{x3} & b_3-\frac{2}{3}a_3 & C_3 \\
         f_{y3} & C_3 &   -b_3-\frac{2}{3}a_3 
    \end{array}
    \right]\,,    
\end{equation}
\begin{equation}
    \label{eq:Jmu_expansion_app}
    \langle J_\mu\rangle =( a_{1t}\,,\,a_{1x}+\partial_xa_t-\partial_ta_x\,,\,a_{1y}+\partial_ya_t-\partial_ta_y)\,,
\end{equation}
\begin{equation}\label{eq:O_expansion}
    \langle \mathcal{O}\rangle = \Psi_{R2}+i\Psi_{I2}\,.
\end{equation}
Note that scalar field does not contribute to $\langle T^{\mu\nu}\rangle$ directly because we turn off the scalar source $\Psi_1$. Nevertheless, presence of scalar hair would affect $\langle T^{\mu\nu}\rangle$ indirectly through bulk gravitational dynamics. Applying~\eqref{eq:a3t}-~\eqref{eq:a1t}, it's straightforward to verify the following Ward identities are satisfied:
\begin{align}
    \nabla_\mu \langle T^{\mu\nu}\rangle &= F^{\mu\nu}\langle J_\mu\rangle\,,\\
     \nabla_\mu \langle J^{\mu}\rangle&=0\,.
\end{align}
In our simulations, the external field strength $F_{\mu\nu}\equiv \partial_\mu a_\nu-\partial_\nu a_\mu=0$ identically since $a_\mu$ is constant throughout the simulations.

\section{Hydrodynamic variables from constitutive relation}

With $\langle T^{\mu\nu}\rangle$, $\langle J^{\mu}\rangle$, and $\langle \mathcal{O}\rangle$ at hand, we can now extract the hydrodynamic variables. For notational simplicity, we omit the angle brackets hereafter. We begin with superfluid velocity $\xi_\mu=\partial_\mu\theta-a_\mu$. We already know $a_\mu$ as boundary conditions for $A_\mu$. $\partial_\mu\theta$ can be deferred from the condensate $\mathcal{O}=|\mathcal{O}|e^{i\theta}$ and its derivative $\partial_\mu\mathcal{O}$:
\begin{align}
    \partial_\mu\theta=-i\frac{\mathcal{\bar O}\partial_\mu\mathcal{O}-\mathcal{O}\partial_\mu\mathcal{\bar O}}{2|\mathcal{O}|^2}\,.
\end{align}
In practice we use $\xi_\mu = (\partial_\mu \theta - a_\mu)\,|\mathcal{O}|/\max|\mathcal{O}|$ to regulate the singularity of $\xi_\mu$ at vortex cores.
For a vortex-free relativistic superfluid, the ideal constitutive relations are
\begin{equation}
T^{\mu\nu}=(\epsilon+p)u^\mu u^\nu+p\eta^{\mu\nu}+\alpha\,\xi^\mu\xi^\nu,
\end{equation}
\begin{equation}
J^\mu=(\rho-\rho_s)u^\mu+\alpha\,\xi^\mu,
\end{equation}
\begin{equation}
u^\mu u_\mu=-1,
\qquad
u^\mu\xi_\mu+\mu=0,
\qquad
\alpha=\frac{\rho_s}{\mu}.
\end{equation}
where $u^\mu$ is the covariant velocity field of normal fluid, $\epsilon$ is the energy density, $p$ is the pressure, $\rho$ is the total charge density (number density), $\rho_s$ is the superfluid denisity and $\mu$ is the chemical potential.
From Eq.~\eqref{eq:Tmunu_expansion} we see $T^\mu_\mu=0$, and therefore $p=(\epsilon-\alpha \xi^2)/2$.
Given the holographic one-point functions $T^{\mu\nu}$, $J^\mu$, and the superfluid velocity $\xi_\mu$, the hydrodynamic variables can be extracted from the constitutive relations. It is convenient to define
\begin{equation}
Q^\mu\equiv J^\mu-\alpha\,\xi^\mu,
\,\,\,
S^{\mu\nu}\equiv T^{\mu\nu}-\alpha\,\xi^\mu\xi^\nu.
\end{equation}
The constitutive relations then reduce to
\begin{equation}
Q^\mu=(\rho-\rho_s)u^\mu,
\,\,\,
S^{\mu\nu}=\frac{3\epsilon-\alpha\xi^2}{2}\,u^\mu u^\nu+\frac{\epsilon-\alpha\xi^2}{2}\eta^{\mu\nu}.
\end{equation}
The normalization condition $u^\mu u_\mu=-1$ implies $(\rho-\rho_s)^2=-Q^2$,
so that the normal-fluid velocity is
\begin{equation}
u^\mu=\frac{Q^\mu}{\sqrt{-Q^2}}.
\end{equation}
Substituting this expression into the stress tensor gives
\begin{equation}
S^{\mu\nu}=\frac{\epsilon-\alpha\xi^2}{2}\eta^{\mu\nu}-\frac{3\epsilon-\alpha\xi^2}{2}\frac{Q^\mu Q^\nu}{Q^2}.
\label{eq:S_perfect}
\end{equation}
The energy density $\epsilon$ and parameter $\alpha$ are determined by requiring that Eq.~\eqref{eq:S_perfect} be satisfied. 
The chemical potential is fixed by the Josephson relation $\mu=-u^\mu\xi_\mu$, and the superfluid density is $\rho_s=\alpha\mu$.
Finally, the normal-fluid density is $\rho_n=\rho-\rho_s=\sqrt{-Q^2}$, so that the total charge density is $\rho=\rho_s+\rho_n$.  

\begin{figure}[htbp]
\centering
\includegraphics[width=0.6\textwidth]{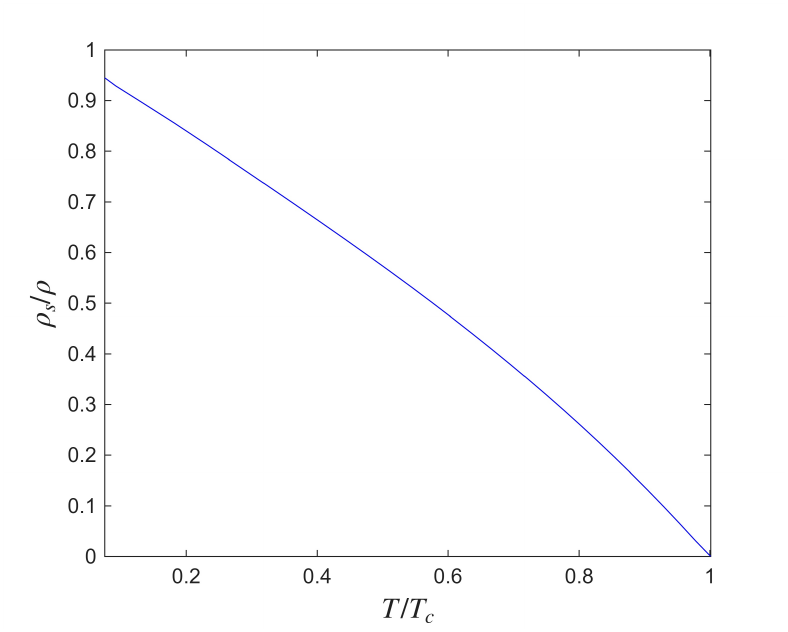}
\caption{Superfluid fraction $\rho_s/\rho$ versus temperature $T/T_c$. Below the critical temperature $T_c$, the superfluid fraction increases as temperature decreases, and tends to 1 at zero temperature, which we can not reach exactly in numerics.
\label{fig:rhos}}
\end{figure}

In Fig.~\ref{fig:rhos}, we present variation of extracted superfluid fraction $\rho_s/\rho$ with respect to temperature for equilibrium states with zero superfluid velocity. Below the critical temperature $T_c$, the superfluid fraction increases as temperature decreases, and tends to 1 at zero temperature, which we can not reach exactly in numerics. Note that for a nonequilibrium evolution, hydrodynamic variables extracted in this way do not correspond to any equilibrium state. This is only leading order approximation in expansion of derivatives. 

\subsection{Considering derivative corrections}\label{app:renormalization}

For a vortex-free conformal superfluid, the general constitutive relations are
\begin{equation}
T^{\mu\nu}=(\epsilon+p)u^\mu u^\nu+p\eta^{\mu\nu}+\alpha\,\xi^\mu\xi^\nu+\pi^{\mu\nu},
\end{equation}
\begin{equation}
J^\mu=(\rho-\rho_s)u^\mu+\alpha\,\xi^\mu+J^\mu_{diss},
\end{equation}
with
\begin{equation}
u^\mu u_\mu=-1,
\qquad
u^\mu\xi_\mu+\mu+\mu_{diss}=0,
\qquad
\alpha=\frac{\rho_s}{\mu}\,,
\end{equation}
where $\pi^{\mu\nu}$, $J^\mu_{diss}$ and $\mu_{diss}$ are contributions from higher order derivatives. In general, one needs to choose a frame to define fields out of equilibrium, including normal fluid velocity $u^\mu$~\cite{bhattacharya2011dissipative}. For example one can choose to work in the frame determined by 
\begin{align}
    &\mu_{diss}=0\,,\quad J^\mu_{diss}=0\,,
\end{align}
and then work out derivative corrections to constitutive relation order by order together with all the transport coefficients. To do this, we need additional input from equilibrium states, i.e., equations of state
\begin{equation}
    \epsilon(\frac{T}{\mu},\frac{\xi}{\mu})\,,\quad p(\frac{T}{\mu},\frac{\xi}{\mu})\,,\quad \rho(\frac{T}{\mu},\frac{\xi}{\mu})\,,\quad \rho_s(\frac{T}{\mu},\frac{\xi}{\mu})\,,
\end{equation}
with $\xi=\sqrt{\xi^\mu\xi_\mu}$. Our aim here is to analysis hydrodynamic variables, so we do not need the explicit form or value of derivative corrections. For this purpose unknown variables are just $(T,\mu)$. Note that temperature $T$ of different frame choice is generally different and does not necessarily agree with local temperature defined in~\eqref{eq:T}. The two equations for determining these two parameters can be chosen as
\begin{align}
    &u^\mu u_\mu=\frac{(J^\mu-\alpha\xi^\mu)^2}{(\rho(\frac{T}{\mu},\frac{\xi}{\mu})-\rho_s(\frac{T}{\mu},\frac{\xi}{\mu}))^2}=-1\,,\\
  & u^\mu\xi_\mu=\frac{(J^\mu-\alpha\xi^\mu)\xi_\mu}{\rho(\frac{T}{\mu},\frac{\xi}{\mu})-\rho_s(\frac{T}{\mu},\frac{\xi}{\mu})}=-\mu\,.
\end{align}
Once we have $(T,\mu)$, we can extract other fields from equations of state and then normal fluid velocity is given by
\begin{equation}\label{eq:u2}
    u^\mu=\frac{J^\mu-\alpha\xi^\mu}{\rho-\rho_s}\,.
\end{equation}

However, even when derivative corrections are included, the hydrodynamic description remains only an approximation in the presence of vortices, as it is expected to break down when the system is far from equilibrium, especially in the vicinity of vortex cores. In such regions, hydrodynamic variables are not well-defined at a fundamental level. The underlying microscopic dynamical variables of the boundary field theory are $T^{\mu\nu}$, $J^\mu$ and $\mathcal{O}$. Nevertheless, in order to compare with experimental observations, it is still necessary to introduce an effective notion of normal fluid velocity, even in vortex-rich configurations.
Since incorporating derivative corrections does not necessarily lead to a more reliable description in the presence of vortices, while significantly increasing computational complexity, in the main text we restrict ourselves to the leading-order hydrodynamic approximation, and extract the normal fluid velocity using Eq.~\eqref{eq:u_reconstruction}.

\section{Freely decaying turbulence}
\label{app:decay}

In this section, we present the extended data and specific parameters for the freely decaying turbulence simulations. The system is initialized at $\bar T/T_c=0.716$ and $\bar\xi_x=0.162$, above Landau critical velocity. Following a small initial spatial perturbation, the system evolves through the instability phase into a decaying turbulent state. 

\begin{figure}[htbp]
\centering
\includegraphics[width=0.6\textwidth]{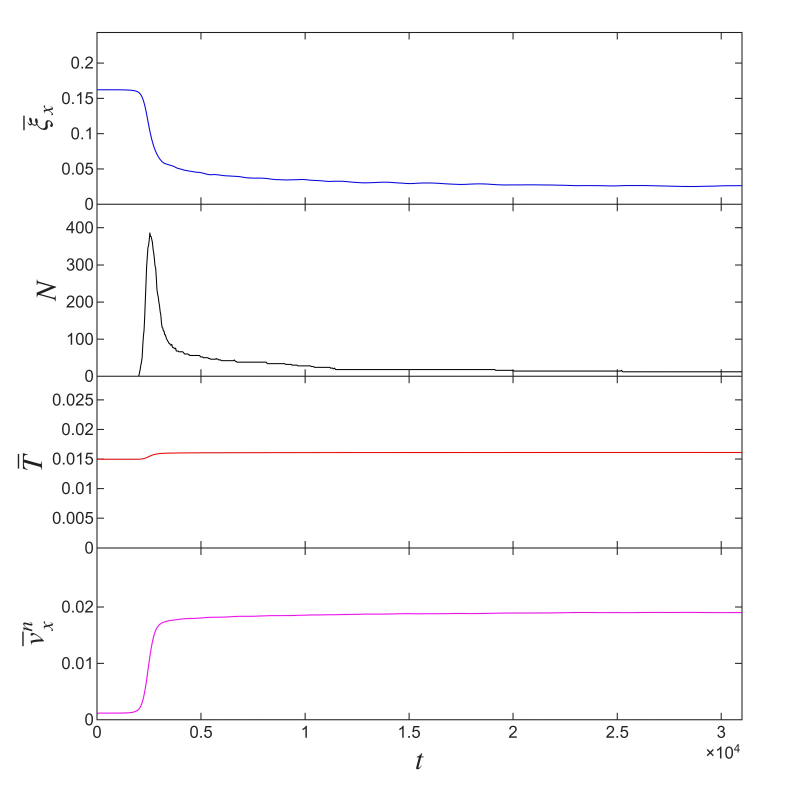}
\caption{Time evolution of the spatially averaged physical quantities during freely decaying turbulence, demonstrating the transition from the unstable superflow to a relaxed sub-critical state. From top to bottom, the panels display the mean superfluid velocity $\bar{\xi}_x$, the total vortex number $N_v$, the mean effective temperature $\bar T$, and the induced mean normal fluid velocity $\bar{v}^n_x$. The concurrent decrease in both the superfluid velocity and the vortex number signifies the decaying nature of the turbulence. Meanwhile, the normal fluid velocity increases as it is dragged by the quantized vortices via mutual friction, and the overall temperature remains nearly constant after the initial vortex nucleation phase.
\label{fig:tevol_decay}}
\end{figure}

Fig.~\ref{fig:tevol_decay} illustrates the time evolution of the spatially averaged physical quantities. While the vortex number and superfluid velocity decay over time, the normal fluid is distinctly dragged along immediately following vortex nucleation due to mutual friction. 

\begin{figure}[htbp]
\centering
\includegraphics[width=0.85\textwidth]{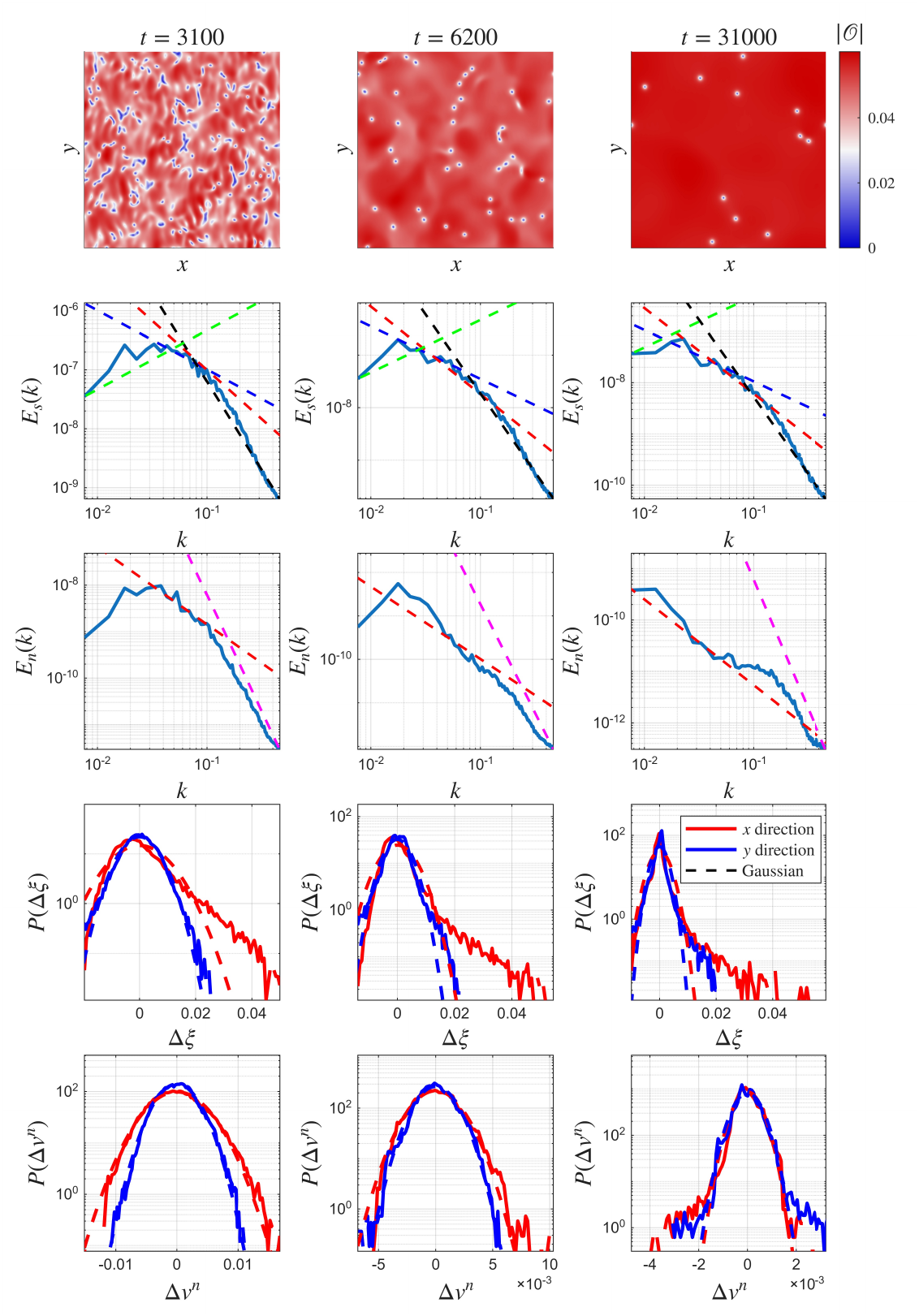}
\caption{\textbf{Snapshots of freely decaying turbulence.} Left to right columns show results at $t=3100$, $6200$, and $31000$. Top to bottom rows plot the order parameter amplitude $|\mathcal{O}|$, superfluid energy spectrum $E_s(k)$, normal fluid energy spectrum $E_n(k)$, and the velocity probability density functions (PDFs) for the superfluid $P(\Delta \xi)$ and normal fluid $P(\Delta v^n)$. For the energy spectra, dashed lines represent power-law scalings: red (Kolmogorov's $k^{-5/3}$), blue ($k^{-1}$), black ($k^{-3}$), green ($k$), and magenta ($k^{-5}$). For the velocity PDFs, colors distinguish spatial directions, while dashed lines show Gaussian fits. The standard deviations $(\sigma_x, \sigma_y)$ for the superfluid and normal fluid velocities are, respectively: $(2.7, 1.8)\times10^{-2}$ and $(4.0, 2.8)\times10^{-3}$ at $t=3100$; $(1.6, 1.2)\times10^{-2}$ and $(1.8, 1.5)\times10^{-3}$ at $t=6200$; $(9.5, 7.5)\times10^{-3}$ and $(0.46, 0.49)\times10^{-3}$ at $t=31000$. It is evident that the normal fluid velocity fluctuations decrease over time, in tandem with the decaying vortex number.
\label{fig:snapshot_decay}}
\end{figure}

\begin{figure}[htbp]
\centering
\includegraphics[width=0.8\textwidth]{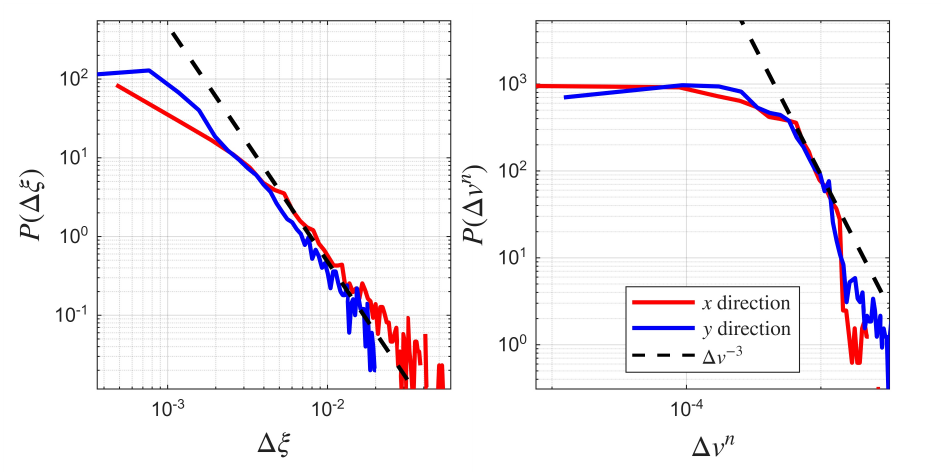}
\caption{The log-log plot of velocity PDFs for the superfluid $P(\Delta \xi)$ and normal fluid $P(\Delta v^n)$ at $t=31000$. A power law tail $\Delta v^{-3}$, which corresponds to single vortex velocity statistics, is observed in both superfluid and normal fluid velocity PDFs.
\label{fig:velocity_statistics}}
\end{figure}

We focus our detailed spatial and statistical analysis on three distinct temporal snapshots during the decay process: $t=3100$, $t=6200$, and $t=31000$. Results are displayed in Figs.~\ref{fig:snapshot_decay}.
Tracking these statistics over time reveals the following stage-specific dynamics:

\begin{enumerate}
    \item As the freely decaying turbulence progresses and the mean flow velocity decreases, the flow-aligned anisotropy in the velocity distributions visibly diminishes. 
    \item The standard deviation of the normal fluid velocity decreases alongside the vortex number, while the mean temperature remains approximately constant following the initial nucleation phase.
    \item By the late stage ($t=31000$), a power law tail $\Delta v^{-3}$, which corresponds to single vortex velocity statistics, is observed in both superfluid and normal fluid velocity PDFs (Fig.~\ref{fig:velocity_statistics}), consistent with previous results in experiments~\cite{paoletti2008velocity} and simulations~\cite{white2010nonclassical,adachi2011numerical,baggaley2011quantum}.
\end{enumerate}

To further elucidate the thermodynamic nature of our setup, it is instructive to contrast the temperature evolution in the freely decaying regime with that of the continuously driven regime. In freely decaying turbulence, the holographic bulk effectively acts as a strictly closed, isolated system. As vortices nucleate and subsequently annihilate, the irreversible dissipation of turbulent kinetic energy naturally leads to a gradual, monotonic increase in the system's temperature due to spontaneous entropy production. Conversely, in the driven scenario, the temperature exhibits a distinctly different trajectory: it slightly decreases during the initial onset of vortex nucleation, briefly recovers, and ultimately stabilizes around a constant plateau in the statistically steady state. This qualitative difference arises because the continuously driven setup fundamentally constitutes an open system interacting with an external control. Our driving operates as an active feedback mechanism that interacts with the system instantaneously at discrete intervals. During these interactions, it injects highly ordered macroscopic kinetic energy (by incrementing the global superfluid phase gradient) while simultaneously extracting an equivalent amount of thermal energy to strictly preserve total energy conservation. This zero-sum energy exchange effectively pumps entropy out of the system, acting as a macroscopic feedback cooling mechanism. In the long-time limit, this external entropy extraction perfectly balances the heat generated by continuous turbulent dissipation, thereby sustaining a pristine non-equilibrium steady state (NESS) without uncontrolled heating. This macroscopic entropy extraction is conceptually analogous to active feedback cooling and laser cooling techniques employed in modern cold-atom experiments, where continuous interaction with an external coherent field removes thermal entropy to stabilize the system in a highly ordered, low-temperature state.

\section{Driven turbulence}
\label{app:diven}

\begin{figure}[htbp]
\centering
\includegraphics[width=0.74\textwidth]{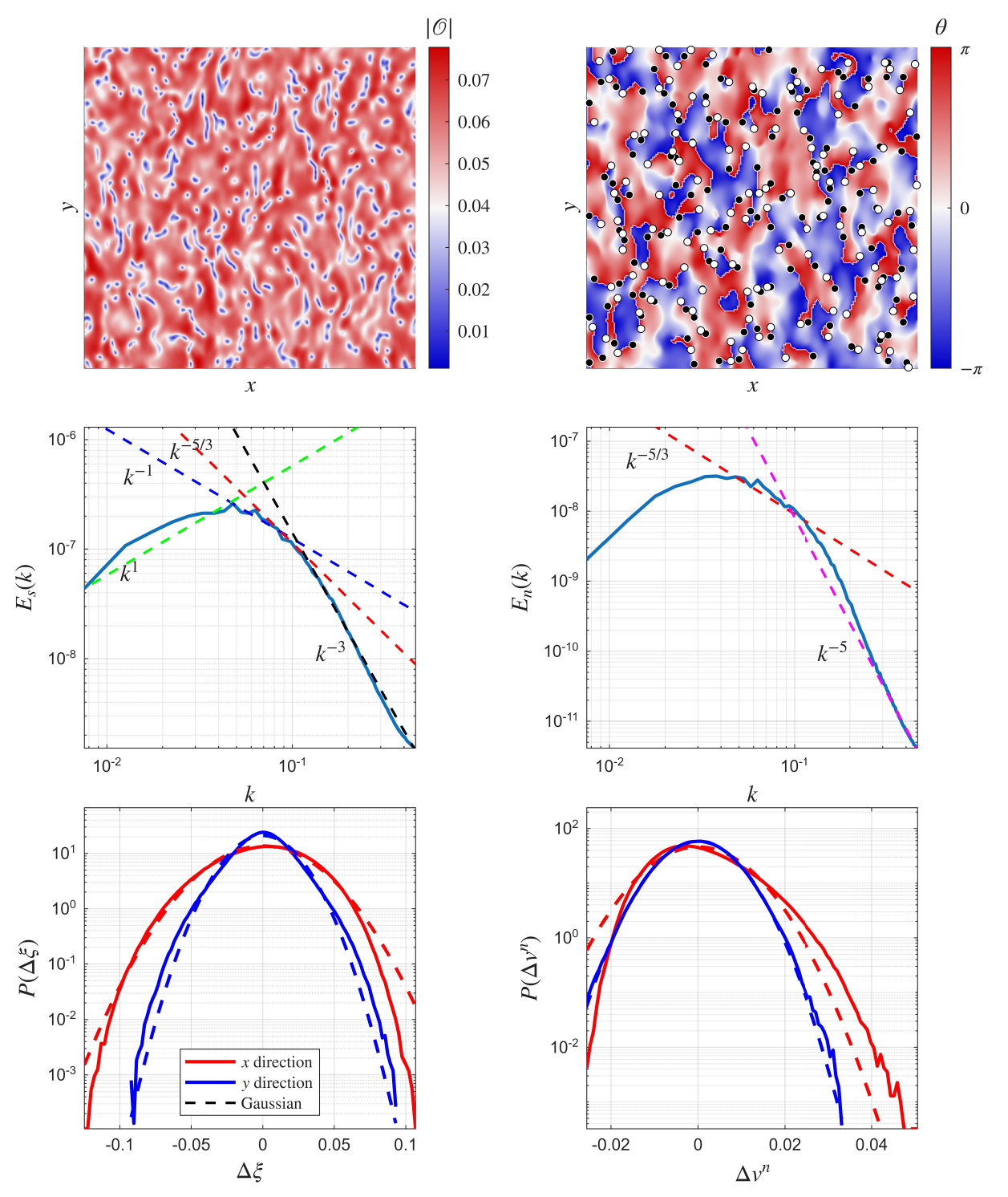}
\caption{\textbf{Statistically steady driven state} at $\bar T/T_c=0.678$ and $\bar\xi_x=0.17$, averaged over 100 time slices in $t\in [6000, 9000]$. \textbf{Top row}: Spatial distribution of the order parameter amplitude $|\mathcal{O}|$ and phase $\theta$ at $t=9000$. White and black dots denote vortices with positive and negative circulations respectively. \textbf{Middle row}: Superfluid $E_s(k)$ and normal fluid $E_n(k)$ energy spectra. Colored dashed lines denote reference power laws: red ($k^{-5/3}$, Kolmogorov), blue ($k^{-1}$), black ($k^{-3}$), green ($k$), and magenta ($k^{-5}$). \textbf{Bottom row}: Velocity probability density functions (PDFs) for the superfluid $P(\Delta \xi)$ and normal fluid $P(\Delta v^n)$, with colors distinguishing spatial directions and dashed lines showing Gaussian fits.
\label{fig:driven2}}
\end{figure}

\begin{figure}[htbp]
\centering
\includegraphics[width=0.74\textwidth]{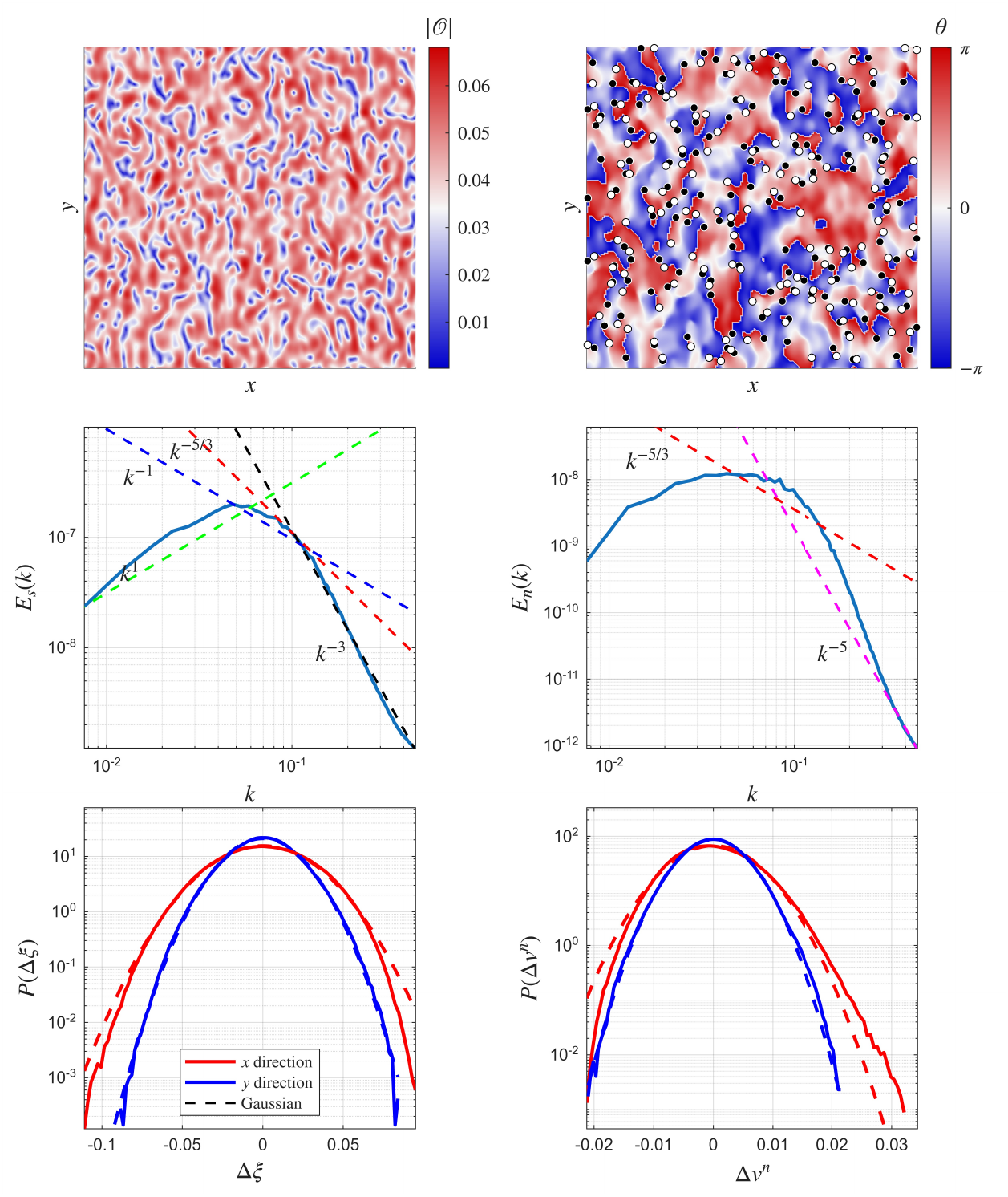}
\caption{\textbf{Statistically steady driven state} at $\bar T/T_c=0.726$ and $\bar\xi_x=0.17$, averaged over 100 time slices in $t\in [6000, 9000]$. \textbf{Top row}: Spatial distribution of the order parameter amplitude $|\mathcal{O}|$ and phase $\theta$ at $t=9000$. White and black dots denote vortices with positive and negative circulations respectively. \textbf{Middle row}: Superfluid $E_s(k)$ and normal fluid $E_n(k)$ energy spectra. Colored dashed lines denote reference power laws: red ($k^{-5/3}$, Kolmogorov), blue ($k^{-1}$), black ($k^{-3}$), green ($k$), and magenta ($k^{-5}$). \textbf{Bottom row}: Velocity probability density functions (PDFs) for the superfluid $P(\Delta \xi)$ and normal fluid $P(\Delta v^n)$, with colors distinguishing spatial directions and dashed lines showing Gaussian fits.
\label{fig:driven3}}
\end{figure}

In this section, we present further visualizations and statistical analyses characterizing the continuously driven turbulent state. Fig.~\ref{fig:driven2} and Fig.~\ref{fig:driven3} display the combined results—comprising real-space snapshots of the order parameter, time-averaged energy spectra, and velocity probability density functions (PDFs)—for a fixed driving velocity of $\bar\xi_x=0.17$ at temperatures $\bar T/T_c=0.678$ and $\bar T/T_c=0.726$, respectively. The order parameter visualizations directly reveal that a higher temperature leads to an increased vortex density in the statistically steady state. The corresponding energy spectra and velocity distributions provide deeper insight into how spatial anisotropy and non-Gaussian statistical features evolve across different temperature regimes. Notably, as the temperature rises, the non-Gaussianity of the normal fluid velocity progressively diminishes, and its energy spectrum deviates from the $k^{-5}$ scaling at large wavenumbers. This behavior is fundamentally attributed to the elevated vortex density, which causes individual vortex cores to crowd and overlap rather than remaining distinctly isolated.

\end{document}